\documentclass[12pt,onecolumn,article]{IEEEtran} 
\usepackage{times,comment}
\usepackage{amsbsy}
\usepackage{latexsym} 
\usepackage{amssymb}
\usepackage{enumitem}
\usepackage{mathtools,xparse}
\usepackage{amsmath,amsfonts,graphicx,epsfig,amsthm,mathtools,diffcoeff}
 \usepackage{times,latexsym,bm,color}
\usepackage[ansinew]{inputenc}
\usepackage{algpseudocode, algorithm}
\usepackage{setspace}
\usepackage{textcomp}
\usepackage{parskip}
\newtheorem{theorem}{Theorem}
\allowdisplaybreaks
\begin{document}
\title{\Large Study of Block Diagonalization Precoding and Power Allocation for Multiple-Antenna Systems with Coarsely Quantized Signals}
\author{Silvio F. B. Pinto and Rodrigo C. de Lamare \vspace{-1.75em} 
\thanks{The authors are with the Center for Telecommunications Studies (CETUC), Pontifical Catholic University of Rio de Janeiro, RJ, Brazil. R. C. de Lamare is also with the Department of Electronics, University of York, UK. Emails: silviof@cetuc.puc-rio.br, delamare@cetuc.puc-rio.br}}
\maketitle 
\begin{abstract} 
In this work, we present block diagonalization and power allocation algorithms for large-scale multiple-antenna systems with coarsely quantized signals. In particular, we develop Coarse Quantization-Aware Block Diagonalization ${\scriptstyle\mathrm{\left(CQA-BD\right)}}$ and Coarse Quantization-Aware Regularized Block Diagonalization ${\scriptstyle\mathrm{\left(CQA-RBD\right)}}$ precoding algorithms that employ the Bussgang decomposition and can mitigate the effects of low-resolution signals and interference. Moreover, we also devise the Coarse Quantization-Aware Most Advantageous Allocation Strategy ${\scriptstyle\mathrm{\left(CQA-MAAS\right)}}$ power allocation algorithm to improve the sum rate of precoders that operate with low-resolution signals. An analysis of the sum-rate performance is carried out along with computational complexity and power consumption studies of the proposed and existing techniques. Simulation results illustrate the performance of the proposed ${\scriptstyle\mathrm{CQA-BD}}$ and ${\scriptstyle\mathrm{CQA-RBD}}$ precoding algorithms, and the proposed ${\scriptstyle\mathrm{CQA-MAAS}}$ power allocation strategy against existing approaches.  
\vspace{-0.7em} 
\end{abstract}

\section{Introduction}
\label{introduction}

{In the last decade, research efforts in wireless communications have shown a great deal of progress in massive multiple-input multiple-output (MIMO) systems. In particular, massive MIMO systems employ base-stations (BS) with a large number of antennas that can serve dozens of users. However, the increasing numbers of antennas at the BS results in higher costs in terms of equipment and power consumption. Therefore, the design of effective and economical massive MIMO systems to equip networks with satisfactory coverage and power consumption will require low-cost components and more energy-efficient algorithms \cite{Rusek,Larsson,Lu,Sarajlic,DeLamare_2013,Wence_cal}. In fact, approaches to energy-efficient design of precoding and detection algorithms have relied on signal quantization with few bits followed by cost-effective strategies such as receive filters, detectors and estimators that can compensate for the loss due to coarse quantization  \cite{Sven1,Sven2,Sven3,Tsinos,Landau,Mezghani}.  Each transmit antenna at the BS is connected to a radio-frequency (RF) chain,
which includes digital-to-analog converters (DACs), low noise amplifier (LNA), mixers, oscillators, automatic gain control (AGC) and filters. Among these components, the power consumption of DACs dominates the total power of the RF chain. In particular, low-resolution DACs are important to reduce the power consumption associated with the transmitter. The power consumption can be substantially reduced when the number of bits used by DACs is reduced because it grows exponentially with the number of quantization bits.}

\subsection{Prior and Related Work}

Despite the progress in 1-bit quantization \cite{Landau, Mezghani} with the aim of reducing power consumption in the large number of DACs used in massive MIMO systems, the achievable sum rates remain relatively low, which makes higher resolution quantizers with $b=2,3,4,\textcolor{red}{5,6}$ bits attractive for the design of linear precoders and receivers. In this context, Bussgang's theorem \cite{Bussgang} let us express Gaussian precoded signals that have been quantized as a linear function of the quantized input and a distortion term which has no correlation with the input \cite{Sven1,Sven2,Sven3}. This approach makes possible the computation of sum-rates of Gaussian signals \cite{Rowe}.

{In particular, block diagonalization ${\scriptstyle\mathrm{\left(BD\right)}}$-type precoding methods \cite{Spencer1, Stankovic, Zu_CL, Zu, Sung,Wence} are known as linear transmit approaches for multiuser MIMO (MU-MIMO) systems based on singular value decompositions (SVD), which provide excellent achievable sum-rates in the case of significant levels of multi-user interference and multiple-antenna users. ${\scriptstyle\mathrm{BD}}$ precoding is motivated by its enhanced sum-rate performance as compared to standard linear zero forcing (ZF) and minimum mean-square error (MMSE) precoders and its suitability for use with power allocation due to the available power loading matrix with the singular values that avoids an extra SVD. However, ${\scriptstyle\mathrm{BD}}$ has not been thoroughly investigated with coarsely quantized signals so far. In addition, existing linear ZF and MMSE precoding techniques that employ 1-bit quantization in massive MU-MIMO systems often present relatively poor performance and significant losses relative to full-resolution precoders. Furthermore, precoding techniques in MU-MIMO systems can greatly benefit from power allocation strategies such as waterfilling. Specifically, power allocation can greatly enhance the sum-rate and error rate performance by employing higher power levels for channels with larger gains and lower power levels for poor channels. Previous works in this area have considered iterative waterfilling techniques \cite{Yu2004}, practical algorihms \cite{Palomar} and specific strategies for ${\scriptstyle\mathrm{BD }}$ precoders \cite{Khan2014} even though there has been no power allocation strategy that takes into account coarse quantization so far, which could enhance the performance of precoders with low-resolution signals.}

\subsection{Contributions}

In this work, we present ${\scriptstyle\mathrm{BD }}$ and power allocation algorithms for large-scale MU-MIMO systems with coarsely quantized signals \cite{cqabd}. Specifically, we develop Coarse Quantization-Aware Block Diagonalization ${\scriptstyle\mathrm{\left(CQA-BD\right)}}$ and Coarse Quantization-Aware Regularized Block Diagonalization ${\scriptstyle\mathrm{\left(CQA-RBD\right)}}$ precoding algorithms that employ the Bussgang decomposition and can mitigate the effects of low-resolution signals and interference. {Moreover, we also devise a Coarse Quantization-Aware Most Advantageous Allocation Strategy ${\scriptstyle\mathrm{\left(CQA-MAAS\right)}}$ power allocation algorithm, which aims to perform the most advantageous power allocation in the presence of coarsely-quantized signals and imperfect channel knowledge to maximize the sum-rate performance.} An analysis of the sum-rate is developed along with a computational complexity study of the proposed and existing techniques. Numerical results illustrate the excellent performance of the proposed ${\scriptstyle\mathrm{CQA-BD}}$ and ${\scriptstyle\mathrm{CQA-RBD}}$ precoding and ${\scriptstyle\mathrm{CQA-MAAS}}$ power allocation algorithms against existing approaches. The main contributions of this work can be summarized as:

\begin{itemize}
    \item {We present the ${\scriptstyle\mathrm{CQA-BD}}$ and ${\scriptstyle\mathrm{CQA-RBD}}$ precoding algorithms for large-scale MU-MIMO systems with coarsely quantized signals.}
    \item{We develop the ${\scriptstyle\mathrm{CQA-MAAS}}$ power allocation algorithm for linearly-precoded MU-MIMO systems.}
    \item{An analysis of the sum-rate is devised along with studies of computational complexity and power consumption.}
    \item{A comparative study of the proposed and existing precoding and power allocation.}
    
\end{itemize}

This paper is structured as follows. Section \ref{sysmodel}
describes the system model and background for understanding the  proposed ${\scriptstyle\mathrm{CQA-BD}}$ class algorithms. Section \ref{proposed_CQA_algorithms} presents the proposed ${\scriptstyle\mathrm{CQA-BD}}$ type algorithms. Section IV introduces the proposed ${\scriptstyle\mathrm{CQA-MAAS}}$ power allocation algorithm, whereas Section V details how precoding and power allocation work together. Section VI analyzes the sum-rate performance, the computational complexity and the power consumption of the proposed algorithms. Section \ref{numerical_results} presents and discusses numerical results whereas the conclusions are drawn in Section \ref{conclusions}.

\textit{Notation}: {the superscript \textit{H} denotes the Hermitian transposition, the superscript \textit{*} stands for the complex conjugate,  $\mathbb E[\cdot]$ expresses the expectation
operator, $\bm I_M$ stands for the $M\times M$  identity matrix, and $\mathbf{0}_{M}$ represents a $M\times 1$ vector whose elements are all zero.}
\section{System Model and Background}
\label{sysmodel}
\begin{figure}[ht] 
	\centering 
	\includegraphics[width=8.3cm, height=4.3cm]{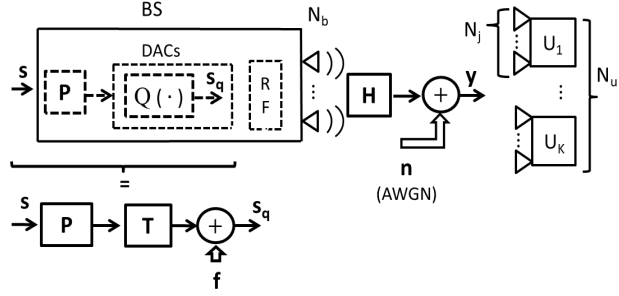} 
	\vspace{-1.0em}
	\caption{Outline of a quantized massive MU-MIMO downlink system. Upper diagram: some simplified parts of BS. Lower diagram: Bussgang's theorem applied to the detached part of interest.}
	\label{sysmodel_CQA_BD}
\end{figure}

Let us consider the broadcast channel (BC) of a MU-MIMO system with a BS containing $N_{b}$ antennas, which sends radio frequency (RF) signals to users equipped with a total of $N_{u}=\sum_{j=1}^{K}\: N_{j}$ receive antennas, where $N_{j}\geq 1$ denotes the number of receive antennas of the $j$th user $U_{j}$, $ j=1,\ldots,K $, as outlined in Fig. \ref{sysmodel_CQA_BD}. 
\vskip-1.5ex
We can model the input-output relation of the BC as
\begin{equation}
\label{downlink_channel_model}
\mathbf{y}= \mathbf{H}\;\mathbf{s}_{q}\:+\:\mathbf{n},
\end{equation}
where $\mathbf{y} \in \mathbb{C}^{N_{u}} $ contains the signals received by all users and $\mathbf{H} \in \mathbb{C}^{N_{u} \times N_{b}} $ stands for the matrix which models the assumed broadcast channel that is assumed known to the BS. The entries of $\mathbf{H}$ are considered independent circularly-symmetrical complex Gaussian random variables $ \left[ \mathbf{H}\right]_{u,b}\in \mathbb{CN} \left( 0,1\right) $, $u=1,\cdots, N_{u}$ and $b=1,\cdots, N_{b}$. The noise vector $\mathbf{n} \in\mathbb{C}^{N_{u}} $  is characterized by its independent and identically distributed (i.i.d.)  circularly-symmetric complex Gaussian entries $n_{u}\in \mathbb{CN} \left( 0,N_{0}\right)$. We consider that the noise variance is known at the BS and so is the sampling rate of  DACs at BS and ADCs at user equipments. 
{In order to provide a better understanding, we include here a short overview of Bussgang's theorem, which allows to deal successfully with nonlinearities like the distortion  generated by DACs.}
{\begin{theorem}
\label{theorem_Busg}
Given two Gaussian signals, the cross-correlation function taken after one of them has undergone nonlinear amplitude distortion is identical, except for a scaling factor, to the cross-correlation function taken before the
distortion \cite{Rowe,Bussgang}. Specifically, according to Bussgang's theorem, for a pair of zero-mean jointly
complex Gaussian random variables $\mathit{y}_{m}\sim \in \mathbb{CN} \left( 0,\sigma_{ym}^{2}\right)$ and $\mathit{y}_{n}\sim \in \mathbb{CN} \left( 0,\sigma_{yn}^{2}\right)$, and for the output $\mathit{r}_{m}$ of some scalar-valued nonlinear function
$\mathit{r}_{m}=  \mathit{f}\left(\mathit{y}_{m}\right)$, where  $\mathit{f}\left(\cdot\right): \mathbb{C}\rightarrow  \mathbb{C}$ expresses an element-wise function application, 
it holds that
\begin{align}
\label{Busg_theo1}
 \mathbb {E}_\mathit{r_{m},y_{n}}\left[\mathit{r_{m}y_{n}}^{*}\right]= \mathit{g_{m}}\mathbb {E}_\mathit{y_{m},y_{n}}\left[\mathit{y_{m}y_{n}}^{*}\right]
\end{align}
in which
\begin{align}
\label{Busg_theo2}
 \mathit{g_{m}}=\frac{1}{\sigma_{ym}^{2}}\mathbb {E}_\mathit{y_{m}}\left[\mathit{f}\left(\mathit{y_{m}}\right)\mathit{y_{m}}^{*}\right]
\end{align}
\end{theorem} }
{
Let us consider now that the  previously mentioned nonlinear function $\mathit{f}\left(\cdot\right)$ can be applied element-wise to a zero-mean complex Gaussian random vector $\mathbf{y}=\left[\mathit{y}_{1},\mathit{y}_{2},\dots, \mathit{y}_{M}\right] \sim \mathbb{CN} \left(\mathbf{0}_{M \times 1}, \mathbf{C_{y}} \right)$, resulting in a vector $\mathbf{f}$, i.e., $\mathbf{r}=  \mathit{f}\left(\mathbf{y}\right)$. It follows from \eqref{Busg_theo1} that}
{
\begin{align}
\label{Busg_theo3}
 \mathbf{C_{ry}}=\mathbf{G\;C_{y}}
\end{align}}
{where
\begin{align}
\label{Busg_theo4}
 \mathbf{G}=\mathrm{diag}\left(\left[\mathit{g}_{1},\mathit{g}_{2},\dots, \mathit{g}_{M} \right]^{T}\right)
\end{align}}
{represents a diagonal $\mathit{M \times M}$ matrix whose \textit{m}th diagonal entry is computed as in \eqref{Busg_theo2}.
Bussgang's theorem can be used to decompose the output of a nonlinear device as a
linear function of the input $\mathbf{y}$ plus a distortion $\mathbf{d}\in \mathbb{C}^{M}$  that is uncorrelated (but not independent) with the input as \cite{Rowe}:}
{
\begin{align}
\label{Busg_theo5}
 \mathbf{r}= \mathbf{Gy }+ \mathbf{d}
\end{align}}
{The referred uncorrelation can be viewed as follows:}
{
\begin{align}
\label{Busg_theo6}
 \mathbb{E}\left[\mathbf{d}\mathbf{y}^{H}\right]&= \mathbb{E}\left[\left(\mathbf{r}-\mathbf{Gy}\right)\mathbf{y}^{H}\right]\nonumber\\&= \mathbf{C_{ry}}-\mathbf{G\;C_{y}}= \mathbf{0}_{\mathit{M} \times \mathit{M} }
\end{align}
where we made use of \eqref{Busg_theo3}.}

Thus, {following the lower part of Fig.\ref{sysmodel_CQA_BD}}, the quantization $\mathrm{Q} \left( \cdot\right)$ of a precoded symbol vector $\mathbf{Ps}$, where $\mathbf{P}\in \mathbb{C}^{Nb\times Nu}$ is a precoding matrix and $\mathbf{s} \sim \mathbb{CN}\left( \mathbf{0}_{Nu\times1},\mathbf{I}_{Nu} \right) $ is the symbol vector, can be expressed by the quantized vector given by
\begin{equation}
\label{bus_theo1}
\mathbf{s}_{q}=\mathrm{Q}\left(\mathbf{Ps}\right)= \mathbf{TPs+f},
\end{equation} 
where the distortion term $\mathbf{f}$, also known as quantization error, and the symbol $\mathbf{s}$ vectors are uncorrelated as shown in \eqref{Busg_theo6} . {Although this term is not Gaussian, in Subsection \ref{subsec_approximations}, we will see that for approximations of achievable sum-rates involving $N_{b}$ and $N_{u}$  sufficiently large, it can be approximated as Gaussian noise, i,e., $\mathbf{f} \in\mathbb{C}^{N_{b}} $, whose entries $f_{b}\in \mathbb{CN} \left( 0,\sigma_{d}^{2}\right) $ }. For the general case, $\mathbf{T}\in \mathbb{R}^{N_{b}\times N_{b}}$ is the diagonal 
matrix expressed {\cite{Sven1}} by 
\begin{align}
\label{diag_mat_busg}
\mathbf{T}_{n,n}=& \frac{\alpha\gamma}{\sqrt{\pi}}\mathrm{diag}\left(\mathbf{P P}^{H} \right)^{-1/2}.\nonumber\\& \sum_{l=1}^{J-1} \exp\left(-\gamma^{2}\left( l-\frac{J}{2}\right)^{2} \mathrm{diag}\left(\mathbf{P P}^{H} \right)^{-1}  \right)  
\end{align} 
where  $n=1\ldots N_{b} $, and $\mathit{J}$ and $\gamma$ stand for the number of levels and the step size of the quantizer, respectively. The scalar factor $\alpha \in \mathbb{R}$, which will be detailed in Subsection \ref{subsec_approximations}, has the purpose of satisfying the average power constraint 
\begin{equation}
\label{power_constraint}
\mathbb {E}\left[\parallel\mathbf {Ps}\parallel_{2}^{2}\right]\leq \mathit{P}
\end{equation}
where the transmit power is given by
\begin{align}
\label{power_x_snr}
 \mathit{P}= \mathrm{SNR} \: \ N_{0},   
\end{align}
where $\mathrm{SNR}$ is the signal-to-noise ratio.

\section{Proposed CQA-BD and CQA-RBD Precoding Algorithms}
\label{proposed_CQA_algorithms}
%

%

In this section, we present the proposed coarse-quantization aware ${\scriptstyle\mathrm{\left(CQA\right)}}$ precoding techniques. In particular, the proposed precoders encompass the ${\scriptstyle\mathrm{CQA-BD}}$ and its regularized version ${\scriptstyle\mathrm{CQA-RBD}}$. The derivation of the ${\scriptstyle\mathrm{CQA-BD}}$ and ${\scriptstyle\mathrm{CQA-RBD}}$ precoding algorithms exploits the knowledge of the channel matrix that contains the channel coefficients of the links between the BS and the users along with SVD operations. In particular, ${\scriptstyle\mathrm{BD}}$ type precoding techniques \cite{Spencer1,Stankovic,Zu} are SVD-based transmit processing algorithms which are performed in two stages. The precoder computed in the first stage suppresses ${\scriptstyle\mathrm{\left(BD \right)}}$ or attemps to obtain a trade-off between MUI and noise ${\scriptstyle\mathrm{\left(RBD\right)}}$. Afterwards, parallel or near-parallel single user (SU)-MIMO are calculated. {The proposed ${\scriptstyle\mathrm{CQA-BD}}$ and ${\scriptstyle\mathrm{CQA-RBD}}$  algorithms can be obtained from the minimization of globally optimum cost functions, which lead to unique solutions to these optimization problems.}

{Both ${\scriptstyle\mathrm{CQA-BD}}$ and ${\scriptstyle\mathrm{CQA-RBD}}$ algorithms compute a precoding matrix $ \mathbf{P}_{j}$ for the \textit{j}th user that can be expressed as the product 
 \begin{equation}
 \label{precod_as_product}
 \mathbf{P}_{j}=\mathbf{P}_{j}^{c}\mathbf{P}_{j}^{d} 
 \end{equation}  
 where $ \mathbf{P}_{j}^{c} \in \mathbb{C}^{N_{b} \times L_{j}}$ and $ \mathbf{P}_{j}^{d} \in \mathbb{C}^{L_{j} \times N_{j}}$. The parameter $L_{j}$ depends on which precoding algorithm is chosen, namely, the ${\scriptstyle\mathrm{ CQA-BD}}$ or ${\scriptstyle\mathrm{CQA-RBD}}$ techniques.}
 
{We can express the combined channel matrix $ \mathbf{H} $ and the resulting precoding matrix $ \mathbf{P} $ as follows:
\begin{equation}
\label{comb_ch_matrix}
 \mathbf{H}=\left[\mathbf{H}_{1}^{T} \mathbf{H}_{2}^{T}\cdots\mathbf{H}_{K}^{T}\right]^{T}\:\in \mathbb{C}^{N_{u} \times N_{b}}
\end{equation}
\begin{equation}
\label{conj_precod_matrix}
\mathbf{P}=\left[\mathbf{P}_{1} \mathbf{P}_{2}\cdots\mathbf{P}_{K}\right]\:\in \mathbb{C}^{N_{b} \times N_{u}}
\end{equation}
where $ \mathbf{H}_{j} \in \mathbb{C}^{N_{j} \times N_{b}} $ is the channel matrix of the $j$th user. The matrix $\mathbf{P}_{j} \in \mathbb{C}^{N_{b} \times N_{j}} $ represents the precoding matrix of the $j$th user.}

\subsection{CQA-BD Precoder}

{In the proposed ${\scriptstyle\mathrm{CQA-BD}}$ precoding algorithm, the first factor  in \eqref{precod_as_product} is given by 
\begin{equation}
\label{precod_mat_first_BD}
\mathbf{P}_{j}^{c\left(CQA-BD \right) }=\overline{\mathbf {W}}_{j}^{\left( 0\right) }
\end{equation}
where $ \overline{\mathbf{W}}_{j}^{\left(0 \right) } $ is obtained by the SVD \cite{Zu} of  \eqref{comb_ch_matrix}, in which the channel matrix of the $j$th  user has been removed, i.e.:
\begin{align}
\label{channel_matrix_user_exclusion}
\overline{\mathbf{H}}_{j}&=\left[\mathbf{H}_{1}^{T}\cdots \mathbf{H}_{j-1}^{T}\mathbf{H}_{j+1}^{T}\cdots\mathbf{H}_{K}^{T}\right]^{T}\:\in \mathbb{C}^{\overline{N}_{j} \times N_{b}}\nonumber\\&= \overline{\mathbf{U}}_{j}\overline{\mathbf{\Phi}}_{j}\overline{\mathbf{W}}_{j}^{H}=\overline{\mathbf{U}}_{j}\overline{\mathbf{\Phi}}_{j}\left[\overline{\mathbf{W}}_{j}^{\left( 1\right) }\overline{\mathbf{W}}_{j}^{\left(0 \right) } \right]^{H} 
\end{align}
where $\overline{N}_{j} =N_{u}-N_{j} $. 
The matrix $ \overline{\mathbf {W}}_{j}^{\left( 0\right) }
 \in \mathbb{C}^{{N}_{b} 
	\times \left( N_{b}-\overline{L}_{j}\right)}$, where $\overline{L}_{j}$ is the rank of $\overline{\mathbf{H}}_{j}$, uses the last $ N_{b}-\overline{L}_{j} $ singular vectors.}

{The second precoder of ${\scriptstyle\mathrm{CQA-BD}}$ in \eqref{precod_as_product} is obtained by SVD  of the effective channel matrix for the $j$th user $\mathbf{H}_{e_{j}}$ and employs a power loading matrix as follows:
\begin{align}
\label{second_factor_BD}
\mathbf{P}_{j}^{d\left(CQA-BD \right) )}=\mathbf{W}_{j}^{\left(1 \right)}\:{\left(\mathbf{\Omega}^{\left(CQA-BD \right)}_{j}\right)^\frac{1}{2}} 
\end{align}
where the power loading matrix $\mathbf{\Omega}^{\left(CQA-BD \right)}_{j}$ requires a power allocation algorithm and the matrix $\mathbf{W}_{j}^{\left(1 \right)}$ incorporates the first $\Lambda_{e}=rank\left( \mathbf{H}_{e_{j}}\right)$ singular vectors obtained by the decomposition of $ \mathbf{H}_{e_{j}} $, as follows: 
\begin{align}
\label{effect_channel_matrix}
\mathbf{H}_{e_{j}}&= \mathbf{H}_{j} \mathbf{P}_{j}^{c}\nonumber= \mathbf{U}_{j}\mathbf{\Phi}_{j}\mathbf{W}_{j}^{H}\\&=\mathbf{U}_{j}
\begin{bmatrix}
\mathbf{\Phi}_{j} & 0\\ 0 & 0
\end{bmatrix}
\begin{bmatrix}
\mathbf{W}_{j}^{\left(1 \right)}&\mathbf{W}_{j}^{\left(0 \right)}
\end{bmatrix}^{H}
\end{align}}
\subsection{CQA-RBD Precoder}

{In the case of the proposed ${\scriptstyle\mathrm{CQA-RBD}}$ precoding algorithm, the global optimization that leads to its design is given by
\begin{equation}
\mathbf{P}_{j}^{c (CQA-RBD)}=\min_{\mathbf{P}_j^c} \mathbb{E}\left[\lVert\bar{\mathbf{H}}_k\mathbf{P}_{j}^c\rVert^2+\frac{\lVert \mathbf{n}_j\rVert^2}{P}\right],\label{CQA-RBD optimization}
\end{equation}}
The first precoder in \eqref{precod_as_product} is given \cite{Stankovic,Zu} by
\begin{align}
\label{precod_mat_first_RBD}
\mathbf{P}_{j}^{c\left(CQA-RBD \right)}= \overline{\mathbf{W}}_{j} \left(\overline{ \mathbf{\Phi}}_{j}^{T} \overline{ \mathbf{\Phi}_{j}} + {\chi}\;\mathbf{I}_{N_{b}}
\right)^{-1/2} 
\end{align}
where $ \chi = \frac{N_{u}\sigma_{n}^{2}}{P} $ is the regularization factor required by the ${\scriptstyle\mathrm{CQA-RBD}}$ algorithm  and $ P $ is the average transmit power.

The second precoder of ${\scriptstyle\mathrm{CQA-RBD}}$ in \eqref{precod_as_product} is obtained by SVD  of the effective channel matrix for the $j$th user $\mathbf{H}_{e_{j}}$ and power loading, respectively as follows:
\begin{align}
\label{second_factor_RBD}
\mathbf{P}_{j}^{d\left(CQA-RBD \right) )} = \mathbf{W}_{j}\:{\left(\mathbf{\Omega}^{\left(CQA-RBD \right)}_{j}\right)^\frac{1}{2}}
\end{align}
where the matrix $\mathbf{W}_{j}^{\left(1 \right)}$ incorporates the early $ \Lambda_{e}=rank\left( \mathbf{H}_{e_{j}}\right)$ singular vectors obtained by the decomposition of $ \mathbf{H}_{e_{j}} $, as follows: 
\begin{align}
\label{effect_channel_matrix}
\mathbf{H}_{e_{j}}&= \mathbf{H}_{j} \mathbf{P}_{j}^{c}\nonumber= \mathbf{U}_{j}\mathbf{\Phi}_{j}\mathbf{W}_{j}^{H}\\&=\mathbf{U}_{j}
\begin{bmatrix}
\mathbf{\Phi}_{j} & 0\\ 0 & 0
\end{bmatrix}
\begin{bmatrix}
\mathbf{W}_{j}^{\left(1 \right)}&\mathbf{W}_{j}^{\left(0 \right)}
\end{bmatrix}^{H}
\end{align}
{The power loading matrix per user {$\mathbf{\Omega}_j^{\left(CQA-RBD \right)}$} can be obtained by a procedure like water filling (WF) \cite{Paulraj} power allocation and will be initialized with equal power allocation}. 

{With Bussgang's decomposition, the transmit processing equivalence $\mathrm{Q}\left(\mathbf{Ps}\right)= \mathbf{TPs+f}$ and the assumption in  \eqref{diag_mat_busg}, we obtain the following transmit processing matrix: 
\begin{equation}
\label{approx_diag_mat0}
\mathbf{T}_{n,n}= \delta\:\mathbf{I}_{Nb\times Nb},
\end{equation} 
where the scalar factor is described by
\begin{equation}
\delta= \alpha \gamma\sqrt{\frac{N_{b}}{\pi P}} \sum_{l=1}^{J-1}\exp\left(-\frac{N_{b}\gamma^{2}}{P}\left( 1-\frac{J}{2} \right)^{2} \right)\, 
\label{entries_diag_mat_dist} 
\end{equation}
which concentrates all process of quantization on the scalar $\delta$ in \eqref{approx_diag_mat0} and is used to compute the sum-rates at the receiver. The loss of achievable sum-rates for a fixed SNR due to the coarse quantization and a fixed realization of the channel are compensated for by $\mathbf{T}_{n,n}$ and $\delta$. The steps needed to compute ${\scriptstyle\mathrm{CQA-BD}}$ and ${\scriptstyle\mathrm{CQA-RBD}}$ are summarized in Algorithm \ref{algorithm:CQA_BD_RBD}. Extensions to other precoders and/or beamforming strategies \cite{wence,1bitcpm,dynovs,sint,sint2,rmmseprec,bapls,rmmsecf,mbthp,rmbthp,rsbd,rsthp,wljio,rdrcb,locsme,okspme,lrcc,armo,baplnc,jpaba} are possible. Moreover, detection and parameter estimation strategies can also be considered for future work \cite{jidf,jio,rrser,spa,mfsic,mbdf,bfidd,1bitidd,1bitadap,1bitce,listmtc}}


\begin{algorithm}[htb!]
	\scriptsize
	\caption{Proposed CQA-BD and CQA-RBD  precoders } \label{algorithm:CQA_BD_RBD}	
	\begin{algorithmic}[1]
		\Require  
	$\begin{aligned} 
		\alpha&= \left( 2\mathrm {N_{b}}\gamma^{2} \left(\left( \frac{J-1}{2}\right)^{2} \right.\right.\nonumber\\&\left.\left. -2\sum_{l=1}^{J-1}  \left( 1-\frac{J}{2} \right)  \Xi \left( \sqrt{2N_{b}\gamma^{2}}\left( 1-\frac{J}{2} \right)\right)\right)\right)^{-1/2} \eqref{normalization_factor}
		\end{aligned} $ 
		$\begin{aligned}
		\delta= \alpha \gamma\sqrt{\frac{N_{b}}{\pi P}} \sum_{l=1}^{J-1}\exp\left(-\frac{N_{b}\gamma^{2}}{P}\left( 1-\frac{J}{2} \right)^{2} \right)\eqref{entries_diag_mat_dist}
		\end{aligned} $ 
         $\begin{aligned}
         \mathbf{H}=\left[\mathbf{H}_{1}^{T} \mathbf{H}_{2}^{T}\cdots\mathbf{H}_{K}^{T}\right]^{T}\:\in \mathbb{C}^{N_{u} \times N_{b}} \:\eqref{comb_ch_matrix}
		\end{aligned} $ 
		\For{$\mathrm{j = 1\;}\colon \: K$}
		\State $\overline{\mathbf{H}}_{j}=\left[\mathbf{H}_{1}^{T}\cdots \mathbf{H}_{j-1}^{T}\mathbf{H}_{j+1}^{T}\cdots\mathbf{H}_{K}^{T}\right]^{T}\:\in \mathbb{C}^{\overline{N}_{j} \times N_{b}}$\eqref{channel_matrix_user_exclusion}
		\State $\overline{\mathbf{H}}_{j}=  \overline{\mathbf{U}}_{j}\overline{\mathbf{\Phi}}_{j}\overline{\mathbf{W}}_{j}^{H}=\overline{\mathbf{U}}_{j}\overline{\mathbf{\Phi}}_{j}\left[\overline{\mathbf{W}}_{j}^{\left( 1\right) }\overline{\mathbf{W}}_{j}^{\left(0 \right) } \right]^{H}$ \eqref{channel_matrix_user_exclusion}
		\State $\mathbf{P}_{j}^{c\left(CQA-BD \right) }=\overline{\mathbf {W}}_{j}^{\left( 0\right) }$\eqref{precod_mat_first_BD}
		\State $\mathbf{P}_{j}^{c\left(RBD \right) }=\overline{\mathbf{W}}_{j}\left( \overline{\mathbf{\Phi}}_{j}^{T} \overline{\mathbf{\Phi}}_{j} + {\chi}\: \mathbf{I}_{N_{b}} \right)^{-1/2} $\eqref{precod_mat_first_RBD}
		\State $ 
		\mathbf{H}_{e_{j}}= \mathbf{H}_{j} \mathbf{P}_{j}^{c}= \mathbf{U}_{j}\mathbf{\Phi}_{j}\mathbf{W}_{j}^{H}=\mathbf{U}_{j} \begin{bmatrix}
		\mathbf{\Phi}_{j} & 0\\ 0 & 0
		\end{bmatrix}
		\begin{bmatrix}
		\mathbf{W}_{j}^{\left(1 \right)}&\mathbf{W}_{j}^{\left(0 \right)}
		\end{bmatrix}^{H}\nonumber $\eqref{effect_channel_matrix}
		
		\State {$\left(\mathbf{\Omega}^{\left(CQA-BD,CQA-RBD \right)}_{j}\right)^\frac{1}{2} $}  \text{by classical WF \cite{Paulraj} or variations} 
		
		\State $
		\mathbf{P}_{j}^{d\left(CQA-BD \right) )}=\mathbf{W}_{j}^{\left(1 \right)}\:{\left(\mathbf{\Omega}^{\left(BD \right)}_{j}\right)^\frac{1}{2}} $ \eqref{second_factor_BD}
		\State $
		\mathbf{P}_{j}^{d\left(CQA-RBD \right) )}=\mathbf{W}_{j}\:{\left(\mathbf{\Omega}^{\left(CQA-RBD \right)}_{j}\right)^\frac{1}{2}} $ \eqref{second_factor_RBD}
		\State $\mathbf{P}_{j}=\mathbf{P}_{j}^{c}\mathbf{P}_{j}^{d}$\:\eqref{precod_as_product}
		
		\EndFor
		\State $ \mathbf{P}=\left[\mathbf{P}_{1} \mathbf{P}_{2}\cdots\mathbf{P}_{K}\right]^{T}\:\in \mathbb{C}^{N_{b} \times N_{u}}  $ \eqref{conj_precod_matrix}
		
	\end{algorithmic}
\end{algorithm}
\subsection{{Precoding and Power Allocation}}
\label{prec+power_s}

{Here, we detail how the proposed ${\scriptstyle\mathrm{CQA-BD}}$ and ${\scriptstyle\mathrm{CQA-RBD}}$ precoding and ${\scriptstyle\mathrm{CQA-MAAS}}$ power allocation algorithms are carried out prior to data transmission.}

{Let us consider the precoding and power allocation using the  ${\scriptstyle\mathrm{CQA-BD}}$ precoder given by
\begin{equation}
\begin{split}
 \mathbf{P}_{j}^{c\left(CQA-BD \right) }&=\mathbf{P}_{j}^{c}\mathbf{P}_{j}^{d} \\
 & =\overline{\mathbf {W}}_{j}^{\left( 0\right)} \mathbf{W}_{j}^{\left(1 \right)}\:{\left(\mathbf{\Omega}^{\left(CQA-BD \right)}_{j}\right)^\frac{1}{2}}
 \end{split}
\end{equation}  
where the power loading matrix $\mathbf{\Omega}^{\left(CQA-BD \right)}_{j}$ is computed according to the ${\scriptstyle\mathrm{CQA-MAAS}}$ power allocation detailed in Section \ref{power_allocation}. In particular, the ${\scriptstyle\mathrm{CQA-BD}}$ precoder (or ${\scriptstyle\mathrm{CQA-RBD}}$ precoder) is computed first with uniform power allocation and then the ${\scriptstyle\mathrm{CQA-MAAS}}$ power allocation is carried out.}

\section{Proposed CQA-MAAS Power Allocation}
\label{power_allocation}

In this section, we derive the
proposed ${\scriptstyle\mathrm{CQA-MAAS}}$  power loading algorithm to compute the matrix $\mathbf{\Gamma}^{\left(CQA-BD\right)} $ based on the waterfilling principle \cite{Cover}. Similar procedure can be pursued for obtaining $\mathbf{\Gamma}^{\left(CQA-RBD\right)}$ and other linear precoders. {In contrast to the design of the proposed ${\scriptstyle\mathrm{CQA-BD}}$ and ${\scriptstyle\mathrm{CQA-RBD}}$ precoders, the proposed ${\scriptstyle\mathrm{CQA-MAAS}}$ power allocation does not lead to a unique solution and is only guaranteed to converge to a local optimum.} Before starting the derivation, we consider the following essential properties and one theorem to facilitate its exposition: 

  \begin{enumerate}[label=(\roman*)]
     \item  Let  $\mathbf{A}$ be a matrix $\in$ $\mathbb{C}^{m \times n} $, $m <n$, with rank $\mathit{r}\leq \mathit{p}=\min \{m,n\}$ \cite{Seber}.
   \begin{enumerate}  
     \item \label{prop_1_a} It can be written as a product $\mathbf{S Y T }^{H}$, which is its SVD.
     \item \label{prop_1_b} $\mathbf{S}$ and $\mathbf{T}$ have orthonormal columns, i.e., $\mathbf{S}^{H}\mathbf{S}=\mathbf{I}_{m}$ and $\mathbf{T}^{H}\mathbf{T}= \mathbf{I}_{n}$.
     \item $\mathbf{Y}$\label{prop_1_c}  has nonnegative elements on its main diagonal and zeros elsewhere.
    \end{enumerate} 
     \item Assuming that $\mathit{k, k_{1}, k_{2}} $ are scalars and the sizes of the matrices $\mathbf{A}$,
     $\mathbf{B} $, $\mathbf{C}$ and $\mathbf{E}$ are chosen so that each operation is well defined, we have \cite{Anton}:
     \begin{enumerate}
    \item $
   \mathit{k}\:\mathbf{AB}=\left(\mathit{k}\mathbf{A}\right)\mathbf{B}= \mathbf{A} \left(\mathit{k}\mathbf{B}\right)$
  \item
  $ \mathbf{A\left(B+C\right)}= \mathbf{AB}+\mathbf{AC}$
  \item $\mathbf{\left(B+C\right)\mathbf{A}}= \mathbf{BA}+\mathbf{CA}$
  \item \label{prop_2_d}
  $\mathit{k}_{1}\:\mathbf{\left(ABC\right)}+\mathit{k}_{2}\:\mathbf{\left(AEC\right)}=\mathbf{A}\left[\left(\mathit{k}_{1}\:\mathbf{B} +\mathit{k}_{2}\:\mathbf{E} \right)\:\mathbf{C}\right] $
  \end{enumerate}
  \item \label{prop_3}Let  $\mathbf{F}$ and $\mathbf{G}$ be $\mathit{m\times n}$ and $\mathit{n\times m}$ matrices, respectively. Then, the following identity holds: $\det\left( \mathbf{I}_{m} +\mathbf{FG}  \right)=$ $\det\left( \mathbf{I}_{n} +\mathbf{GF}  \right)$ \cite{Telatar}.
  \item \label{prop_4} 
Let $\mathbf{A}= \left[ \mathrm{a}_{ij} \right]$ be a non-negative $ n \times n$ Hermitian matrix. Then,
$ \det\left(\mathbf{A}\right) \leq \prod_{i=1}^{n} \mathrm{a}_{ii}$ 
with equality if and only if some $\mathrm{a}_{ii}=0$  or $\mathbf{A}$ is diagonal \cite{Telatar}.
\item \label{prop_5}
    $\mathrm{\log \det\left(\mathbf{A}\right)}=\mathrm{Tr\log\left(\mathbf{A}\right),\:}\mathrm{\forall}\mathrm{\:nonsingular\:} \mathbf{A}^{n\times n} \text{\cite{Withers}} $
  \item \label{prop_6}
  $\det\left(\mathbf{A}\right)=\prod_{i=1}^{n}\:  \lambda_{i}$ where $\lambda_{i}$ denotes the $\mathit{i}$th eigenvalue of  $\mathbf{A} \:\in\mathbb{C}^{n \times n}$   or $\mathbb{R}^{n \times n} $\cite{Seber}.
  \item \label{prop_7}
  $\log_{2}\det\left(\mathbf{A}\right)=\sum_{i=1}^{n}\log_{2}\left(\mathbf{a}_{ii}\right), \quad \mathrm{if} \:\mathbf{A}\:  \mathrm{is\:diagonal}\\ \mathrm{and} \quad  \mathbf{a}_{ii} >0$
   \end{enumerate}
 
We now resort to the expression of the  achievable sum-rate of ${\scriptstyle\mathrm{CQA}}$ precoding algorithms whose derivation is detailed in the Appendix:    
 \begin{align}
\label{achievable_sum_rate_BD_power_alloc}
\mathit{C}=& \log_{2}\left\lbrace\det\left[ \mathbf{I}_{Nu} + \delta^{2} \mathbf{\left( HP\right) }\mathbf{R_{s}} \mathbf{\left( HP\right) }^{H}\right.\right.\nonumber\\&\left.\left. \left(\left(1-\delta^{2} \right)\mathbf{\left( HP\right) } \mathbf{R_{s}}\mathbf{\left( HP\right) }^{H} +\mathrm{N}_{0}\mathbf{I}_{Nu}              \right)^{-1}   \right]  \right\rbrace 
\end{align}
where $\mathbf{R_{s}=\mathbb E[\mathbf{s}\mathbf{s}^{H}]}$ denotes the covariance matrix associated to the symbol vector. The expression comprises a factor composed of the inverse of a sum of matrices, which is hard to deal with. An alternative approach to computing it would be a recursive procedure \cite{Miller}, which could not lead to a compact form. Since we  aim to obtain closed-form expressions, we have opted for  employing the following approximation based on Neumann's truncated matrix series \cite{Seber,Harville}. 
\begin{theorem}
\label{theorem1}
Let $\mathbf{Q}$ represent an $ n \times n$ matrix. Then, the infinite series $\mathbf{I}+\mathbf{Q}+\mathbf{Q}^{2}+\mathbf{Q}^{3}+\cdots$ converges if and only if  $\lim_{k\to\infty}\mathbf{Q}^{k}=\mathbf{0}$, in which case $\mathbf{I}-\mathbf{Q}$ is non-singular and 
\begin{align}
\label{Newmann}
 \left(\mathbf{I}-\mathbf{Q}\right)^{-1}&= \sum_{k=0}^{\infty}\left(\mathbf{\mathbf{Q}} \right)^{k}\nonumber\\&=\mathbf{I}+\mathbf{Q}+\mathbf{Q}^{2}+\mathbf{Q}^{3}+\cdots 
\end{align}
\end{theorem}
where $\mathbf{Q}^{0}=\mathbf{I} $.
\newline
Replacing $\mathbf{Q}$ by $\mathbf{\left(-Q\right)}$, we have
\begin{align}
\label{Newmann2}
 \left(\mathbf{I}+\mathbf{Q}\right)^{-1}&= \sum_{k=0}^{\infty}\left(\mathbf{\mathbf{-Q}} \right)^{k}\nonumber\\&=\mathbf{I}-\mathbf{Q}+\mathbf{Q}^{2}-\mathbf{Q}^{3}+\cdots 
\end{align}
\begin{theorem}
\label{theorem2}
Let $\mathbf{Q}$ represent an $ n \times n$ matrix. If $\| \mathbf{Q} \|<1$, where $\|\cdot\|$ denotes the Frobenius norm.
Then, $\lim_{k\to\infty}\mathbf{Q}^{k}=\mathbf{0}$
\end{theorem}
Based on Theorem \ref{theorem2}, we assume that  \eqref{Newmann2} can provide accurate approximations if we truncate it after the second term. Then, we  employ 
\begin{align}
\label{Newman_3}
  \left(\mathbf{I}+\mathbf{Q}\right)^{-1} \approx \mathbf{I}-\mathbf{Q}.  
\end{align}
for rewriting \eqref{achievable_sum_rate_BD_power_alloc} as follows: 
\begin{align}
\label{achievable_sum_rate_BD_power_alloc_1}
\mathit{C}=& \log_{2}\left\lbrace\det\left[ \mathbf{I}_{Nu} + \frac{\delta^{2}}{\mathrm{N}_{0}} \mathbf{\left( HP\right) }\mathbf{R_{s}} \mathbf{\left( HP\right) }^{H}\right.\right.\nonumber\\&\left.\left.\times \left(\mathbf{I}_{Nu} +\underbrace{ \overbrace{\frac{\left(1-\delta^{2} \right)}{\mathrm{N}_{0}}}^{\epsilon}\mathbf{\left( HP\right) } \mathbf{R_{s}}\mathbf{\left( HP\right) }^{H}}_{{\mathbf{Q}}}  \right)^{-1}   \right]  \right\rbrace \nonumber\\& \approx \log_{2}\left\lbrace\det\left[ \mathbf{I}_{Nu} + \frac{\delta^{2}}{\mathrm{N}_{0}} \mathbf{\left( HP\right) }\mathbf{R_{s}} \mathbf{\left( HP\right) }^{H}\right.\right.\nonumber\\&\left.\left.  - \frac{\delta^{2}\left(1-\delta^{2} \right)}{\mathrm{N}_{0}}\left(\mathbf{\left( HP\right) } \mathbf{R_{s}}\mathbf{\left( HP\right) }^{H}\right)^2     \right]  \right\rbrace
\end{align}
It can be noticed that the coefficient $\epsilon=\frac{\left(1-\delta^{2} \right)}{\mathrm{N}_{0}}$ constrains the norm of ${\mathbf{Q}}$ matrix resulting from the Hermitian matrices product. For the analysis of the conditions and consequences of the previously assumed approximation, which is discussed in Subsection \ref{Maximum_accurate_SNR_analysis}, we define that coefficient as follows:
\begin{align}
\label{coef_epsilon}
 \epsilon=\frac{\mathit{SNR}\left(1-\delta^{2} \right)}{\mathrm{Nu}}
\end{align}
where  we defined $\mathrm{N}_{0}=\frac{\mathrm{Nu}}{\mathit{SNR}}.
$
It is well known \cite{Telatar} that the capacity for MIMO channels, subjected to ergodicity requirements, when $\mathbf{H}$ is perfectly known at the receiver, can be expressed by
\begin{align}
\label{AWGN_channel}
 \mathit{C}= \log_{2}\left\lbrace \det \left[ 
  \mathbf{I}+\frac{\mathbf{H}\:\mathbf{Q}\:\mathbf{H}^{H}}{\sigma_{n}^{2}}\right]\right\rbrace
\end{align}
where $\sigma_{n}^{2}= \mathbb{E}\left\lbrace n\left(i \right ) n^{H}\left(i \right )\right\rbrace $ denotes the noise covariance 
It is also shown  in \cite{Telatar} that \eqref{AWGN_channel} is maximized when $\mathbf{Q}$ is diagonalized, i.e., $\mathbf{Q}=\mathbf{W}{\Omega}\mathbf{W}^{H}$, where $\mathbf{W}$ $\in$ $\mathbb{C}^{Nb\times Nb}$  denotes the unitary matrix containing the eigenvectors of $\mathbf{H}\mathbf{H}^{H}$ described by
\begin{align}
\label{Diag_power_matrix}
 \mathbf{\Omega}= 
 \begin{bmatrix}
 \mathit{\omega}_{11} & 0 & \cdots & 0\\
    0  & \mathit{\omega}_{22} &\cdots & 0 \\
    \vdots & & \ddots & \\
    0 & \cdots & \cdots  & \mathit{\omega}_{NN} 
 \end{bmatrix}
\end{align}
where the total power is distributed among  the diagonal entries of $\mathbf{\Omega}$, i.e.,   $\sum_{i=i}^{N}\left[\mathbf{\Omega}\right]_{ii}=\mathit{P}_{t}$.
For unquantized ${\scriptstyle\mathrm{BD}}$ or ${\scriptstyle\mathrm{RBD}}$ precoding algorithms, i.e., a full resolution one, we can expand \eqref{AWGN_channel} in terms of the $N_{j}$ parallel sub channels obtained from the SVD of the non-interfering block channels  \eqref{effect_channel_matrix} as follows: 
\begin{align}
\label{AWGN_channel_expand}
 \mathit{C}_{fr_j}= &\log_{2}\left\lbrace \det \left[
 \mathbf{I}_{N_j}\right.\right.\nonumber\\\nonumber\\ &\left.\left.+\frac{\mathbf{U}_{j}\:\mathbf{\Phi}_{j}\:\mathbf{W}_{j}^{H}\:\mathbf{W}_{j}\:\mathbf{\Omega }_{j}\:\mathbf{W}_{j}^{H}\: \left(\mathbf{U}_{j}\:\mathbf{\Phi}_{j}\:\mathbf{W}_{j}^{H}\right)^{H}}{\sigma_{n}^{2}}
 \right]\right\rbrace 
\end{align}
 We can use properties \ref{prop_1_b} and \ref{prop_3} to transform \eqref{AWGN_channel_expand} into 
\begin{align}
\label{AWGN_channel_expand_1}
 \mathit{C}_{fr_j}= 
 \log_{2}\left\lbrace \det \left[
 \mathbf{I}+\frac{\mathbf{\Phi}_{j}^2\:\mathbf{\Omega }_{j}}{\sigma_{n}^{2}}
 \right]\right\rbrace 
\end{align}
The total sum-rate for unquantized ${\scriptstyle\mathrm{BD}}$ can be computed as the sum-rates of each user, as follows:
\begin{align}
\label{AWGN_channel_expand2}
 \mathit{C}_{fr}= \sum_{j=1}^{K} \log_{2}
 \left\lbrace \det \left[
  \mathbf{I}_{Nj}+\frac{\mathbf{\Phi}_{j}^2\:\mathbf{\Omega }_{j}}{\sigma_{n}^{2}}
  \right]\right\rbrace 
 \end{align}
 where we can neglect the  $\mathit {\left(N_b-N_u\right) }$ right column vectors of $\mathbf{\Phi}_{j}$
   $\in $ $\mathbb{R}^{Nj \times \left[ Nb-\left(N_u-N_j\right)\right]}$. Thus, we consider only the left main diagonal matrix, as indicated by 
 \begin{align}
 \label{singular_values}
   \mathbf{\Phi}_{j}=
   \underbrace{\begin{bmatrix}
   \phi_{11} & 0 & \cdots & 0  \\
     0        & \phi_{22} & \cdots & 0\\
   \vdots & & \ddots &  \vdots\\
    0 & \cdots & \cdots  & \phi_{N_jN_j}
   \end{bmatrix}}_{ \begin{aligned}&\: Nj\: {\rm valid} \\& {\rm diagonal}\: {\rm matrix} \end{aligned}}
   \underbrace{\left[\begin{matrix}
   0 & \cdots & 0 \\
   0 & \cdots & 0 \\
   \vdots &\cdots &   \vdots\\
   0 & \cdots & 0
   \end{matrix}\right].}_{ \begin{aligned}&\:(N_b-N_u)~{\rm null}\\ &{\rm vectors}\: {\rm neglected} \end{aligned}}
 \end{align}
In order to make the size of the power allocation matrix $ \mathbf{\Omega }_{j}$ described ahead compatible with the size of the singular values  matrix $\mathbf{\Phi}_{j}$, we take into account only  its upper left $\mathit{Nj}$ diagonal matrix, which consists  of the underbraced elements :
\begin{align}
\label{Diag_power_matrix_each_user}
 \mathbf{\Omega}_{j}= 
 \begin{bmatrix}
 \underbrace{{\omega}_{1 1}} &  \bullet\bullet\bullet & 0 & 0  &\cdots&  0\\
    \boldsymbol{\vdots}& \boldsymbol{ \ddots} & \boldsymbol{\vdots}  & \vdots &\ddots &\vdots \\
    0 & \bullet\bullet\bullet & \underbrace{{\omega}_{NjNj}}& 0 &\cdots  & 0 \\
    0 &\cdots& 0  & 0 &\cdots & 0\\
   \vdots &  \ddots & \vdots & \vdots &\ddots &\vdots \\
    0  & \cdots & 0 & 0 & \cdots & 0
 \end{bmatrix}
\end{align}
The problem of power allocation for an unquantized ${\scriptstyle\mathrm{BD}}$ algorithm can then be formulated as follows:
\begin{align}
\label{AWGN_channel_expand2_1}
 \mathit{C}_{fr}= \max_{\Phi_{j}} \sum_{j=1}^{K} \log_{2}
 \left\lbrace \det \left[
  \mathbf{I}_{N_j}+\frac{\mathbf{\Phi}_{j}^2\:\mathbf{\Omega }_{j}}{\sigma_{n}^{2}}\nonumber
  \right]\right\rbrace 
 \end{align}
 \begin{align}
 \mathrm{s.t.} \: \sum_{j=1}^{K} \mathrm{Tr} \left(\mathbf{\Omega}_{j}\right)\leq\:\mathit{P}_{total}  
 \end{align}
Now, we can expand the approximation in \eqref{achievable_sum_rate_BD_power_alloc_1}  by means of the $\mathit{N}_{j}$ parallel sub channels obtained from the SVD of the non-interfering block channels  \eqref{effect_channel_matrix}.  The procedure is  similar to that used to obtain \eqref{AWGN_channel_expand}. For compactness, we use the notation $\det\left( \cdot\right)= \left|\cdot\right|$.  
\begin{align}
\label{AWGN_channel_expand3}
\mathit{C}_{j} &\approx \log_{2}\bigg|
  \mathbf{I}_{N_j}\nonumber\\ & +\frac{\delta^{2}}{\mathrm{N}_{0}}\mathbf{U}_{j}\:\mathbf{\Phi}_{j}\:\mathbf{W}_{j}^{H}\:\mathbf{W}_{j}\:\mathbf{\Omega}_{j}\:\mathbf{W}_{j}^{H}\: \left(\mathbf{U}_{j}\:\mathbf{\Phi}_{j}\:\mathbf{W}_{j}^{H}\right)^{H} \nonumber\\ &-\frac{\delta^{2}\left(1-\delta^{2} \right)}{\mathrm{N}_{0}}\left(\mathbf{U}_{j}\:\mathbf{\Phi}_{j}\:\mathbf{W}_{j}^{H}\:\mathbf{W}_{j}\:\mathbf{\Omega }_{j} \mathbf{W}^{H}_{j} \right.\nonumber\\ & \left.\times \left(\mathbf{U}_{j}\:\mathbf{\Phi}_{j}\:\mathbf{W}_{j}^{H}\right)\right)^{2}
  \bigg|
\end{align}
By using properties \ref{prop_1_b} and \ref{prop_2_d}, expression \eqref{AWGN_channel_expand3} can be simplified as follows:
\begin{align}
\label{AWGN_channel_expand4}
\mathit{C}_{j} &\approx \log_{2}\bigg|
  \mathbf{I}_{N_j} +\mathbf{U}_{j}\left(\frac{\delta^{2}}{\mathrm{N}_{0}}\:\mathbf{\Phi}_{j}^{2}\:\mathbf{\Omega }_{j}\right.\nonumber\\&\left. -\frac{\delta^{2}\left(1-\delta^{2} \right)}{\mathrm{N}_{0}^{2}}\mathbf{\Phi}_{j}^{4}\:\mathbf{\Omega }^{2}\right)\mathbf{U}_{j}^{H}
  \bigg|\nonumber\\ & =\log_{2}\bigg|
  \mathbf{I}_{N_j} +\frac{\delta^{2}}{\mathrm{N}_{0}}\mathbf{\Phi}_{j}^{2}\:\mathbf{\Omega }_{j}\nonumber\\ &-\frac{\delta^{2}\left(1-\delta^{2} \right)}{\mathrm{N}_{0}^{2}}\mathbf{\Phi}_{j}^{4}\:\mathbf{\Omega }^{2}
  \bigg|
\end{align}
The argument of the determinant  is composed of three terms,  which are  diagonal matrices composed of real  entries  and whose  sum provides  a real diagonal vector. This allows us to combine property \ref{prop_4} with property \ref{prop_5} to formulate the maximization needed for power allocation of our proposed ${\scriptstyle\mathrm{CQA-MAAS}}$ power allocation approach. The procedure is similar to that provided for obtaining the sum-rate  for full resolution ${\scriptstyle\mathrm{BD}}$ in \eqref{AWGN_channel_expand2_1}, i.e.:
\begin{align}
\label{form_maximization_CQA_BD}
 \mathit{C}&\approx  \max_{\Phi_{j}} \sum_{j=1}^{K} \log_{2}
 \bigg|
  \mathbf{I}_{N_j}+\frac{\delta^{2}}{\mathrm{N}_{0}}\:\mathbf{\Phi}_{j}^{2}\:\mathbf{\Omega }_{j}  -\frac{\delta^{2}\left(1-\delta^{2} \right)}{\mathrm{N}_{0}^{2}}\:\mathbf{\Phi}_{j}^{4}\:\mathbf{\Omega }_{j}^{2}
  \bigg|\nonumber\\
  & = \max_{\Phi_{j}} \sum_{j=1}^{K} \mathrm{Tr}\left[\mathrm{\log}_{2}
 \bigg|
  \mathbf{I}_{N_j}+\frac{\delta^{2}}{\mathrm{N}_{0}}\:\mathbf{\Phi}_{j}^{2}\:\mathbf{\Omega }_{j}  \right.\nonumber\\& \left.- \frac{\delta^{2}\left(1-\delta^{2} \right)}{\mathrm{N}_{0}^{2}}\:\mathbf{\Phi}_{j}^{4}\:\mathbf{\Omega }_{j}^{2}
  \bigg| \right] \nonumber\\
  &\qquad \mathrm{s.t.}  \sum_{j=1}^{K} \mathrm{Tr} \left(\mathbf{\Omega}_{j}\right)\leq\:\mathit{P}_{total}
  \end{align}
  where $\mathbf{\Phi}_{j}$ and $\mathbf{\Omega}_{j}$ are expressed by \eqref{singular_values} and \eqref{Diag_power_matrix_each_user}, respectively.

 Now, we can deal with the constrained optimization problem \cite{Paulraj,Cover,Palomar} in \eqref{form_maximization_CQA_BD} with the method of Lagrange multipliers. To this end, we write the cost function as
 \begin{align}
 \label{cost_function_CQA_BD}
  \Upsilon \left(\mathbf{\Omega}_{1},\cdots, \mathbf{\Omega}_{K}\right) &= \sum_{j=1}^{K} \mathrm{Tr}\left[\mathrm{\log}_{2}
 \bigg|
  \mathbf{I}_{N_j}+\frac{\delta^{2}}{\mathrm{N}_{0}}\:\mathbf{\Phi}_{j}^{2}\:\mathbf{\Omega }_{j}  \right.\nonumber\\& \left.- \frac{\delta^{2}\left(1-\delta^{2} \right)}{\mathrm{N}_{0}^{2}}\:\mathbf{\Phi}_{j}^{4}\:\mathbf{\Omega }_{j}^{2}
  \bigg| \right]+ \beta\sum_{j=1}^{K} \mathrm{Tr} \left(\mathbf{\Omega}_{j}\right)
 \end{align}
 where $\beta$ denotes the Lagrange multiplier. By using property \ref{prop_7}, we can convert \eqref{cost_function_CQA_BD} into
 \begin{small}
 \begin{align}
 \label{cost_function_CQA_BD_matrix_form}
  \Upsilon \left(\mathbf{\Omega}_{1},\cdots, \mathbf{\Omega}_{K}\right) &= \sum_{j=1}^{K} \mathrm{Tr}
  \begin{bmatrix}
        \log_{2}\left(\mathrm{d}_{11}\right) && \\
        & \ddots&\\
        & & & \log_{2}\left(\mathrm{d}_{nn}\right)
  \end{bmatrix}_{j}
  \nonumber\\& +\beta\sum_{j=1}^{K} \mathrm{Tr}
  \begin{bmatrix}
        \mathrm{\omega}_{11} && \\
        & \ddots&\\
        & & & \mathrm{\omega}_{NjNj},
  \end{bmatrix}_{j}
  \end{align}
  \end{small}
  where the subscript $\mathit{\left(\cdot\right)_{nn}}$ of the diagonal entries is  associated with the $\mathit{n}$th receive antenna of the $\mathit{j}$th user. In the case of the first summand, we have  
 \begin{align}
 \label{diag_entries_cost}
  \mathrm{d}_{nn}\big|_{n=1, \cdots, N_{j}}=1+\frac{\delta^{2}}{\mathrm{N}_{0}}\:\mathbf{\phi}_{nn}^{2}\:\mathbf{\omega }_{nn}-\frac{\left(\delta^{2}-\delta^{4} \right)}{\mathrm{N}_{0}^{2}}\:\mathbf{\phi}_{nn}^{4}\:\mathbf{\omega }_{nn}^{2}   
 \end{align}
  We can combine \eqref{cost_function_CQA_BD_matrix_form} with \eqref{diag_entries_cost} and maximize the resulting cost function associated with a generic $\mathit{n}$th receive antenna of the $j$th user as follows:
    \begin{align}
  \label{cost_func_each_ant}
   &\diffp{\mathbf{\Upsilon}\left(\omega_{nn}\right)}{\omega_{nn}}=\frac{1}{\ln\left(2\right)}\:\frac{1}{\mathrm{d}_{nn}}\: \diffp{\left(\mathrm{d}_{nn}\right)}{\omega_{nn}} + \beta\: \diffp{\left(\omega_{nn}\right)}{\omega_{nn}}=0
  \end{align}
  By solving \eqref{cost_func_each_ant}, we can obtain the energy level, defined as
 \begin{align}
 \label{ener_level}
  \mu=\frac{-1}{\beta \ln \left(2\right) }=\frac{\mathrm{N}_{0}^{2}+\mathrm{N}_{0}\delta^{2}\phi^{2}_{nn}\omega_{nn}- \left( \delta^{2}-\delta^{4}\right)\phi_{nn}^{4}\omega_{nn}^{2}}{\mathrm{N}_{0}^{2}\delta^{2} \phi^{2}_{nn}- 2\left( \delta^{2}-\delta^{4}\right)\phi_{nn}^{4}\omega_{nn}}
 \end{align}
  \underline{Note}: From this point on, we will drop the subscripts $\left(\cdot\right)_{nn}$  for the sake of compactness.
  
  After algebraic manipulations and rearranging terms in \eqref{ener_level}, we obtain the following second degree equation:
  \begin{small}
  \begin{align}
  \label{gen_power_eq}
    \omega^{2}- \frac{\left[\mathrm{N}_{0}\delta^{2}\phi^{2}+2 \mu \left( \delta^{2}-\delta^{4}\right)\phi^{4}\right]}{\left( \delta^{2}-\delta^{4}\right)\phi^{4}}\;\omega + \frac{\mu\mathrm{N}_{0}\delta^{2}\phi^{2} -\mathrm{N}_{0}^{2}}{\left( \delta^{2}-\delta^{4}\right)\phi^{4}} =0,
  \end{align}
  \end{small}
  where the distortion factor $\delta$ lies in the range $\left(0 \:1\right)$.
  The solution of \eqref{gen_power_eq} can be arranged as follows:
  \begin{small}
  \begin{align}
  \label{eq_power_CQA}
   \omega= \frac{\mathrm{N}_{0}+2\mu \left( 1-\delta^{2} \right)\phi^{2}}{2 \left( 1-\delta^{2} \right)\phi^{2}}\pm \frac{\sqrt{4\mu^{2}\delta^{2}\left( 1-\delta^{2} \right)^{2}\phi^{4}+\mathrm{N}_{0}^{2}\left(4-3\delta^{2}\right)}}{2 \delta \left( 1-\delta^{2} \right)\phi^{2}},
  \end{align}
  \end{small}
 The procedure of squaring  the second summand, which at a glance could be useful, would imply squared terms of individual powers $\omega^{2}$ that would not satisfy the power constraint \eqref{form_maximization_CQA_BD}, which requires linear terms $ \omega$. Thus, we can express the radical as a McLaurin series truncated after the $\mathit{4}$th term as follows:
  \begin{align}
  \label{approx_radical}
   \mathit{f\left(\phi\right)}=\sqrt{4\mu^{2}\delta^{2}\left( 1-\delta^{2} \right)^{2}\phi^{4}+\mathrm{N}_{0}^{2}\left(4-3\delta^{2}\right)} &\nonumber\\\approx  \sqrt{\mathrm{N}_{0}\left(4-3\delta^{2}\right)}  +  \frac{2\mu^{2}\delta^{2} \left( 1-\delta^{2} \right)\phi^{4}}{\mathrm{N}_{0}^{2}\left(4-3\delta^{2}\right)}
  \end{align}
  where $\mathit{f'\left(\mathrm{0}\right)}=\mathit{f''\left(\mathrm{0}\right)}=\mathit{f'''\left(\mathrm{0}\right))}$
  We can combine \eqref{approx_radical} with \eqref{eq_power_CQA} to obtain the desired power expression:
  \begin{small}
  \begin{align}
  \label{single_power_expression}
   \omega=\frac{\mathrm{N}_{0}}{2  \left( 1-\delta^{2} \right)\phi^{2}} + \mu \pm \left[\frac{\mathrm{N}_{0}\sqrt{\left(4-3\delta^{2}\right)}}{2 \delta \left( 1-\delta^{2} \right)\phi^{2}} +\frac{\mu^{2}\delta \left( 1-\delta^{2} \right)\phi^{2}}{\mathrm{N}_{0}\sqrt{\left(4-3\delta^{2}\right)}}\right]  
  \end{align}
  \end{small}
  In order to ensure real levels of power, we choose the negative sign in \eqref{single_power_expression}  and rearrange it.  In this way, we arrive at the can formulate the ${\scriptstyle\mathrm{CQA-MAAS}}$ power allocation applied to each  antenna of each user:
   \begin{align}
  \label{negat_single_power_expression}
   \omega^{\rm MAAS}_{nn} \bigg\rvert_{j} =\bigg( \mathit{C}_{1}\mathrm{N}_{0}\frac{1}{\phi^{2}} + \mu  -\mu^{2}\mathit{C}_{2}\frac{1}{\mathrm{N}_{0}}\phi^{2} \bigg)^{\left(+\right)} \bigg\rvert_{j}
  \end{align}
  \begin{align}
   \label{c1}  
   \mathit{C}_{1}=\frac{\delta -\sqrt{4-3\delta^{2}}}
   {2\delta \left( 1-\delta^{2} \right)}
  \end{align}
  \begin{align}
   \label{c2} 
  \mathit{C}_{2}=\frac{\delta\left(1-\delta^{2} \right)}
   {\sqrt{4-3\delta^{2}}}
  \end{align}
  where $\left(+\right)$ denotes the Kuhn-Tucker conditions: 
    \begin{eqnarray}
    \left(\omega_{nn}\right)^{+}=\omega, &  \mathrm{if} \: \omega_{nn}\geq 0 \\
        0, & \mathrm{if} \: \omega_{nn}< 0
  \label{kt_condit}
  \end{eqnarray}
   Now, we can write the constraint in \eqref{form_maximization_CQA_BD} in matrix form as
  \begin{align}
  \label{matrix_form_constraint}
   \sum_{j=1}^{K} \mathrm{Tr} \left(\Omega_{j} \right)=& \mathit{P}_{total}=\sum_{j=1}^{K}
    \begin{bmatrix}
   \mathrm{\omega}_{11} && \\
        & \ddots&\\
        & & & \mathrm{\omega}_{NjNj}\nonumber 
  \end{bmatrix}_{j}
  \\
   =& \sum_{j=1}^{K} \mathrm{Tr}
   \begin{Bmatrix}\left( \mathit{C}_{1}\mathrm{N}_{0}\right)
   \begin{bmatrix}
   \mathrm{\frac{1}{\phi_{11}^{2}}} && \\
        & \ddots&\\
        & & & \mathrm{\frac{1}{\phi^{2}_{\mathit{NjNj}}}}\nonumber 
  \end{bmatrix}_{j}
  \end{Bmatrix}
  \\
   +& \sum_{j=1}^{K} 
   \mathrm{Tr}
  \begin{Bmatrix}
   \mu
  \begin{bmatrix}
   1 && \\
        & \ddots&\\
        & & & 1_{NjNj}\nonumber 
  \end{bmatrix}_{j}
  \end{Bmatrix}
  \\
   -& \sum_{j=1}^{K} \mathrm{Tr}
   \begin{Bmatrix}
   \mu^{2}\left( \frac{\mathit{C}_{2}}{\mathrm{N}_{0}}\right)
  \begin{bmatrix}
   \mathrm{\phi_{11}^{2}} && \\
        & \ddots&\\
        & & & \mathrm{\phi^{2}_{\mathit{NjNj}}}
  \end{bmatrix}_{j}
  \end{Bmatrix}
  \end{align}
After rearranging terms, the matrix-form constraint \eqref{matrix_form_constraint} can be written as a  second degree  equation whose addends are in summation-form:
  \begin{align}
  \label{second_degree_eq}
   \mu^{2}\left( \frac{\mathit{C}_{2}}{\mathrm{N}_{0}}\right)\sum_{j=1}^{K}\sum_{l=1}^{Nj}\left[\phi^{2}\right]_{l,j} - \mu \sum_{j=1}^{K}\sum_{l=1}^{Nj}\left[1\right]_{l,j} \nonumber\\+\mathit{P}_{t} -\left( \mathit{C}_{1}\mathrm{N}_{0}\right)\sum_{j=1}^{K}\sum_{l=1}^{Nj}\left[\frac{1}{\phi^{2}}\right]_{l,j} =0
  \end{align}
The solutions of equation  \eqref{second_degree_eq} can be expressed as:
  \begin{align}
  \label{power_level_solution}
    \mu&=\frac{\mathit{N}_{u}\mathrm{N}_{0}}{2\mathit{C}_{2}} \left(\frac{1}{\sum_{j=1}^{K}\sum_{l=1}^{Nj}\left[\phi^{2}\right]_{l,j}}\right)
    \nonumber
    \\ &\pm \left[\left(\frac{\mathit{N}_{u}\mathrm{N}_{0}}{2\mathit{C}_{2}} \left(\frac{1}{\sum_{j=1}^{K}\sum_{l=1}^{Nj}\left[\phi^{2}\right]_{l,j}}\right)\right)^2  \right.\nonumber\\& \left.   - \left(\frac{\mathit{P}_{t}\mathrm{N}_{0}-\mathit{C}_{1}\mathrm{N}_{0}^{2}\sum_{j=1}^{K}\sum_{l=1}^{Nj}\left[\frac{1}{\phi^{2}}\right]_{l,j}}{\mathit{C}_{2}\sum_{j=1}^{K}\sum_{l=1}^{Nj}\left[\phi^{2}\right]_{l,j} }\right)  \right]^{\frac{1}{2}}
   \end{align}
   We can simplify the previous expression  by considering \eqref{power_x_snr} and recalling, from Section \ref{sysmodel}, that $\mathbf{R}_{ss}\approx \mathbf{I}_{N_{u}} $. We can also assume that the total power  is given by $\mathit{P}= {\rm trace}\left( \mathbf{R}_{ss}\right)= \mathrm{N}_{u} $, resulting in
   $\mathrm{N}_{0}=\frac{\mathit{N}_{u}}{\mathrm{SNR}}$. The combination of this expression with \eqref{power_level_solution} yields:
   \begin{align}
    \label{power_level_solution_ref}
  \mu_{opt}&=\frac{\mathit{N}_{u}^{2}}{2\mathit{C}_{2}\mathrm{SNR}\sum_{j=1}^{K}\sum_{l=1}^{Nj}\left[\phi^{2}\right]_{l,j}}\nonumber \\&\times \Bigg\{ 1 - \left[ 1 + \frac{4\mathit{C}_{2}}{\mathit{N}_{u}^{2}}\sum_{j=1}^{K}\sum_{l=1}^{Nj}\left[\phi^{2}\right]_{l,j} \right.\nonumber\\& \times \left. \left( -\mathrm{SNR}\:+ \mathit{C}_{1} \sum_{j=1}^{K}\sum_{l=1}^{Nj}\left[\frac{1}{\phi^{2}}\right]_{l,j}\right)                \right]^{\frac{1}{2}}\Bigg\}
  \end{align}
  where we  choose the negative sign before the square root in \eqref{power_level_solution}  to ensure the appropriate smallest positive levels of power.
 In summary, the following steps can describe the iterative process of the ${\scriptstyle\mathrm{CQA-MAAS}}$ for power allocation, which is a customization of the classical Waterfilling method \cite{Cover,Paulraj,Palomar} for low-resolution signals. The procedure is derived from expressions \eqref{power_level_solution_ref} and \eqref{negat_single_power_expression}.
 We start with the first one \eqref{power_level_solution_ref}, which can be converted into:
   \begin{align}
   \label{power_level_solution_implem}
  \mu_{opt}&=\frac{\left(\mathit{N}_{u}-\mathit{p}+1\right)^{2}}{2\mathit{C}_{2}\mathrm{SNR}\sum_{m=1}^{\left(\mathit{N}_{u}-\mathit{p}+1\right)}\left[\phi^{2}\right]_{m}}\nonumber \\&\times \Bigg\{ 1 - \left[ 1 + \frac{4\mathit{C}_{2}}{\mathit{N}_{u}^{2}}\sum_{m=1}^{\left(\mathit{N}_{u}-\mathit{p}+1\right)}\left[\phi^{2}\right]_{m} \right.\nonumber\\& \times \left. \left( -\mathrm{SNR}\:+ \mathit{C}_{1} \sum_{m=1}^{\left(\mathit{N}_{u}-\mathit{p}+1\right)}\left[\frac{1}{\phi^{2}}\right]_{m}\right)                                     \right]^{\frac{1}{2}}\Bigg\},
   \end{align}
   where
   \begin{enumerate}[label=(\roman*)]
   \vspace{-2pt}
       \item \label{implem_1}$\mathit{N}_{u}$ stands for the number of receive antennas defined in Section \ref{sysmodel}.
       \item \label{implem_2}$\mathit{p} $ denotes an auxiliary parameter to be set to $1$.
       \item \label{implem_3}$\mathit{C}_{1}$ and $\mathit{C}_{2}$, which depend only on the distortion factor $\delta$, are given  by \eqref{c1} and \eqref{c2}.
       \item \label{implem_4} $\phi$ designates each of the $\mathrm{N}_{u}=\mathit{K \times N_{j}}$ singular values corresponding to each receive antenna. This can be better visualized in  the following diagonal matrix \eqref{Diag_power_matrix}, in which the diagonal vector displays the required entries $\phi_{m}$ $\in$ $\phi_{1}, \cdots,\phi_{\mathit{N_{u}}}$  .
 \begin{scriptsize}
 \begin{align}
\label{Diag_sv_matrix}
 \mathbf{\Phi}= 
 \begin{bmatrix}
 \phi_{1} &  \cdots & 0 &\cdots\cdots & 0  &\cdots&  0\\
    \vdots&  \ddots & \vdots & & \vdots &\ddots &\vdots \\
    0 & \cdots & \phi_{Nj}& \cdots\cdots &0 &\cdots  & 0 \\ \vdots & & \vdots &\ddots &\vdots & &\vdots\\\vdots & & \vdots& &\vdots & &\vdots\\0 &  \cdots & 0 &\cdots\cdots & \phi_{\left(\mathrm{N_{u}}-Nj\right)}  &\cdots&  0\\
    \vdots&  \ddots & \vdots & & \vdots &\ddots &\vdots \\
    0 & \cdots & 0 & \cdots\cdots &0 &\cdots  & \phi_{\mathit{N_{u}}}
 \end{bmatrix}
\end{align}
\end{scriptsize}
 \end{enumerate}
Employing the value of $\mu_{opt}$ provided by \eqref{power_level_solution_implem}, the power allocated to the $\mathit{m}$th $\in$ $\{1, \cdots,\mathrm{N_{u}}\} $ receive antenna can be computed by
   \begin{align}
  \label{single_power_expression_for_alg}
   \omega_{m} &= \mathit{C}_{1}\frac{\left(\mathit{N}_{u}-\mathit{p}+1\right)}{\mathrm{SNR}}\:\frac{1}{\phi^{2}_{m}} + \mu_{opt} \nonumber\\  &-\mu^{2}_{opt}\mathit{C}_{2}\frac{\mathrm{SNR}}{\left(\mathit{N}_{u}-\mathit{p}+1\right)}\:\phi^{2}_{m} 
  \end{align}
 where the parameters involved were defined in \ref{implem_1}, \ref{implem_2}, \ref{implem_3} and \ref{implem_4}. 
  Assuming that the power alloted to the receive antenna which is associated to the  minimum gain is negative, i.e., $\omega_{\mathit{N}_{u}-\mathit{p}+1}< 0$, it is rejected, and the algorithm must be executed with the parameter $\mathit{p}$ increased by unity. The 
  most advantageous allotment strategy is achieved at the time that  the power distributed among each receive antenna is non-negative.

  Here, we sum up the CQA-MAAS power allocation algorithm, which aims to compute 
  $\mathbf{\Omega}^{\left(BD \right)}_{j}$\eqref{second_factor_BD}, in Algorithm \eqref{algorithm:Alg_power_allot}. For this purpose we can form  a larger power diagonal matrix, where each of its $\mathrm{Nu}$ entries is  associated to its corresponding $\mathit{j}$th user, in ascending order, as follows:
  \begin{align}
   \label{}
   \mathbf{\Omega}^{\left(BD\right)}= \mathrm{diag}\{\mathbf{\Omega}_{1}, \cdots,\mathbf{\Omega}_{K} \},  
  \end{align}
  which means that
 \begin{align}
\label{Diag_power_matrix_exp}
 \begin{scriptsize}\mathbf{\Omega}^{\left(BD\right)}= 
 \begin{bmatrix}
 \omega_{1} &  \cdots & 0 &\cdots\cdots & 0  &\cdots&  0\\
    \vdots&  \ddots & \vdots & & \vdots &\ddots &\vdots \\
    0 & \underbrace{\cdots}_{\mathbf{\Omega}_{1}} & \omega_{Nj}& \cdots\cdots &0 &\cdots  & 0 \\ \vdots & & \vdots &\ddots &\vdots & &\vdots\\\vdots & & \vdots& &\vdots & &\vdots\\0 &  \cdots & 0 &\cdots\cdots & \omega_{\left(\mathrm{N_{u}}-Nj\right)}  &\cdots&  0\\
    \vdots&  \ddots & \vdots & & \vdots &\ddots &\vdots \\
    0 & \cdots & 0 & \cdots\cdots &0 &\underbrace{\cdots}_{\mathbf{\Omega}_{K}}  & \omega_{\mathit{N_{u}=K \times Nj}}
 \end{bmatrix}
 \end{scriptsize}
\end{align}
\begin{algorithm}[htb!]
	 \scriptsize
		\caption{Proposed CQA-MAAS power allocation } \label{algorithm:Alg_power_allot}	
	{\begin{algorithmic}[1]
		\State $ \mathbf{Initialization}: \mathrm{Nu}, \mathrm{Nb}, \mathit{K} $\\
		\begin{scriptsize}
 \begin{align}
 \hspace{-40mm}\mathbf{\Phi}= 
 \begin{bmatrix}
 \phi_{1} &  \cdots & 0 \\
    \vdots&  \ddots & \vdots\\
    0&\cdots&\phi_{\mathit{N_{u}}} \nonumber
 \end{bmatrix}\eqref{Diag_sv_matrix}
\end{align}
\end{scriptsize}
		\State
		$\mathbf{Compute}$ the factors $\mathit{C}_{1}$ in \eqref{c1} and $\mathit{C}_{2}$ in \eqref{c2}. \vspace{0.5em}
		\State
		$\mathbf{Compute}$ the optimum energy level $\mu_{opt} $ in \eqref{power_level_solution_ref}.\vspace{0.5em}
		\State 
		$\mathbf{Compute}$ the power allocation to each sub-channel $\omega_{m} $ in \eqref{single_power_expression_for_alg}.\vspace{0.5em}
		\State 
		 $\mathbf{If}$ there are negative values, then find their minimum, i.e., their $\min \left(\omega_{N_{u}-p+1}<0\right) $ $\mathbf{and}$. \vspace{0.5em}
		 \State 
		 $\mathbf{Refuse}$ this minimum negative value by assuming it is equal to zero in \eqref{kt_condit}, $\mathbf{and}$. \vspace{0.5em}
		 \State
		 $\mathbf{Perform}$ the algorithm with the parameter $\mathit{p}$ incremented by unity. \vspace{0.5em}
		 \State
		 CQA-MAAS achieves its goal when the power allocated among the receive antennas is non-negative \eqref{kt_condit}. \vspace{0.5em}
		 \State
		 $\mathbf{Compute}$ the power diagonal matrix \eqref{Diag_power_matrix_exp}, by relating its $N_{u}$ power entries to their corresponding receive antennas.
		\end{algorithmic}}
\end{algorithm} 
Moreover, according to \cite{Spencer1} \cite{Sung}, we can combine \eqref{Diag_power_matrix_exp} with \eqref{second_factor_BD}, \eqref{precod_as_product}, \eqref{precod_mat_first_BD} and \eqref{conj_precod_matrix}, to incorporate ${\scriptstyle\mathrm{CQA-MAAS}}$ into the precoding matrix, expressed in  terms of each user as follows:
\begin{align}
\label{optimal_precoding}
\mathbf{P}_{opt}^{BD}&=\left[\overline{\mathbf {W}}_{1}^{\left( 0\right) }\mathbf{W}_{1}^{\left(1 \right)}\:\left(\mathbf{\Omega}^{\left(BD \right)}_{1}\right)^\frac{1}{2} \cdots\overline{\mathbf {W}}_{K}^{\left( 0\right) }\mathbf{W}_{K}^{\left(1 \right)}\:\left(\mathbf{\Omega}^{\left(BD \right)}_{K}\right)^\frac{1}{2}\right]\nonumber\\
&=\left[\overline{\mathbf {W}}_{1}^{\left( 0\right) }\mathbf{W}_{1}^{\left(1 \right)} \cdots\overline{\mathbf {W}}_{K}^{\left( 0\right) }\mathbf{W}_{K}^{\left(1 \right)}\right]\:\left(\mathbf{\Omega}^{\left(BD \right)}\right)^{\frac{1}{2}}
\end{align}
%




\section{Analysis}
\label{analysis}

In this section, we analyze aspects of the approximation via Neumann's series, which is employed in the formulation of the proposed ${\scriptstyle\mathrm{CQA-MAAS}}$ power allocation. We also examine
the achievable sum-rate of the proposed ${\scriptstyle\mathrm{CQA}}$ precoding and power allocation techniques along with their computational complexity. Moreover, we assess the power consumption of the proposed and existing approaches.
{
\subsection{Maximum accurate SNR under ${\scriptstyle\mathrm{CQA-MAAS}}$ }
\label{Maximum_accurate_SNR_analysis}
In this section, we estimate the maximum SNR for which the approximation proposed in \eqref{Newman_3} provides accurate values of the sum rate in \eqref{achievable_sum_rate_BD_power_alloc_1}.
In \eqref{achievable_sum_rate_BD_power_alloc_1} and \eqref{coef_epsilon} of Section \ref{power_allocation}, we stated that the parameter $\epsilon$ confines the norm of the Hermitian matrices product, which is denoted by $\mathbf{Q}$. It can be seen in \cite{Seber} that  the norm of a product  between a constant $\mathit{c}$ and a  matrix $\mathbf{Q} \in \mathcal{C}^{m \times n}$ obeys $\|\mathit{c}\; \mathbf{Q}\|= \left|\mathit{c}\right|\:\| \mathbf{Q}\|$, where $\left|\cdot\right|$ stands for  the modulus. Moreover, $\epsilon $ \eqref{coef_epsilon} plays the role of $\mathit{c}$ previously described, which can be viewed as a norm shortener. Thus, we are interested in its small values that satisfy the limitation imposed by the theorem in \ref{theorem2} to validate the approximation in \eqref{achievable_sum_rate_BD_power_alloc_1}. In light of the assumed broadcast system, we focus on two  configurations, $ N_{b}=64 $  and $ N_{u}=8 \times 2 $ and $ N_{b}=64 $  and $ N_{u}=16 \times 2 $, which are examined in the simulations in Section \ref{numerical_results}. We also assume an arbitrary value $\epsilon \leq 0.01$. }

{The maximum SNR for which the approximation proposed in  \eqref{achievable_sum_rate_BD_power_alloc_1} provides accurate values of sum rates can be obtained by
\begin{align}
\frac{\mathrm{SNR}\left(1-\delta^2\right)}{\mathit{N}_{u} }\leq 0.01  
\end{align}
which after manipulations, yields
\begin{align}
\label{SNR_dB_max}
 \mathrm{SNR\left(dB\right)}_{max} = 10\;\log_{10}\frac{0.01 \mathit{N}_{u}}{\left(1-\delta^{2}\right)}
\end{align}
Then, based on \eqref{SNR_dB_max} and the two MU-MIMO configurations mentioned before, we can build the following table:}
\begin{table}[htb]\small
   \caption{{Maximum accurate SNR for two MU-MIMO configurations: 1)$ N_{b}=64 $ and $ N_{u}=8 \times 2 $;\hspace{3mm}  2)$ N_{b}=64$ and $ N_{u}=16 \times 2 $ }}
    \centering
    {
    \begin{tabular}{c|c|c|c}
    \hline
      Quantization    & $\delta$ & $ N_{b}=64 $  & $ N_{b}=64 $  \\
      bits &  & $ N_{u}=8 \times 2 $ & $ N_{u}=16 \times 2 $\\
      \hline
      &&&\\
      2 & 0.9387 & 1.2915  & 4.3018 \\
      3 & 0.9811 & 6.3075  & 9.3178 \\
      4 & 0.9942 & 11.4092 & 14.4195 \\
      5 & 0.9983 & 16.7301 & 19.7404  \\
      6 & 0.9995 & 22.0243 & 25.0525 \\
      \hline
    \end{tabular} }
    \label{Table_4}
\end{table}
{Table \ref{Table_4} can be used to roughly estimate the region of the SNR range from which ${\scriptstyle\mathrm{CQA-BD-MAAS}}$ starts to lose efficiency in terms of achievable sum rates, as will be shown in  Fig.\ref{R_ant_16_T_ant_64_b_2_3_4_5_6_200_2}  and Fig.\ref{R_ant_32_T_ant_64_b_2_3_4_5_6_2}, in Section \ref{numerical_results}.} 
\subsection{Achievable sum-rates}
\label{subsec_approximations}
{It is feasible  to compute approximations of achievable sum-rates for downlink channels in which $N_{b}$ and $N_{u}$ are sufficiently large as both the error resulting from the combination of multiuser interference (MUI) and the quantization error from limited resolution of DACs can be considered as Gaussian \cite{Sven1}. This assumption, which is justified by the central limit theorem enables us to convert \eqref{diag_mat_busg} into the matrix:}
\begin{equation}
\label{approx_diag_mat}
\mathbf{T}_{n,n}= \delta\:\mathbf{I}_{Nb\times Nb}
\end{equation}
where the  entries of $\mathbf{T}_{n,n}$  are given by the Bussgang scalar factor:   
\begin{align}
\label{entries_diag_mat_dist_1}
\delta= \alpha \gamma\sqrt{\frac{N_{b}}{\pi P}}
 \: \sum_{l=1}^{J-1}\exp\left(-\frac{N_{b} \gamma^{2}}{P}\left( 1-\frac{J}{2} \right)^{2} \right)  
\end{align}
{where the factor $\alpha$ is obtained by}
\begin{align}
\label{normalization_factor}
 \alpha=& \left( 2\mathrm {N_{b}}\gamma^{2} \left(\left( \frac{J-1}{2}\right)^{2} \right.\right.\nonumber\\&\left.\left. -2\sum_{l=1}^{J-1}  \left( 1-\frac{J}{2} \right)  \Xi \left( \sqrt{2N_{b}\gamma^{2}}\left( 1-\frac{J}{2} \right)\right)\right)\right)^{-1/2}  
\end{align}
where $\Xi \left( w \right) =\int_{- \infty}^{w} \frac{1}{\sqrt{2 \pi}}\exp^{-v^{2}/2}\:dv$ \cite{Sven1}  is the distributed function of a Gaussian random variable. {We have followed the approach of \cite{Sven1} that considers the same sampling rates at both transmitter and receiver and that the DACs have coarse quantization but the ADCs have infinite quantization in order to focus on the effects of the DACs.}
In \cite{Pinto}[Appendix], assuming the system model in Section \ref{sysmodel}  and  the identity in  \eqref{approx_diag_mat}, we have derived the following closed-form approximation for the   sum rate achieved by ${\scriptstyle\mathrm{CQA-BD}}$ and ${\scriptstyle\mathrm{CQA-RBD}}$ precoders via Bussgang's theorem:
\begin{align}
\label{achievable_sum_rate_BD_RBD}
\mathit{C}=& \log_{2}\left\lbrace\det\left[ \mathbf{I}_{Nu} + \delta^{2} \frac{\mathit{SNR}}{\mathit{N}_{u}}\mathbf{\left( HP\right) } \mathbf{\left( HP\right) }^{H}\right.\right.\nonumber\\&\left.\left. \left(\left(1-\delta^{2} \right)\frac{\mathit{SNR}}{\mathit{N}_{u}}\mathbf{\left( HP\right) } \mathbf{\left( HP\right) }^{H} +\mathbf{I}_{Nu}              \right)^{-1}                 \right]  \right\rbrace 
\end{align}
where the $\mathit{SNR}$ was defined in \eqref{power_constraint} and $\mathbf{P}$ is the precoding matrix \eqref{conj_precod_matrix}, which is defined in Section \ref{proposed_CQA_algorithms}.

Note that the quantization effect is concentrated on the Bussgang's factor $\delta$ in \eqref{approx_diag_mat} and \eqref{entries_diag_mat_dist_1}, which approximates the effects of quantization with few bits.
\subsection{Power consumption and efficient DACs}
\label{increasing_requirement_efficiency}

Until recently, the use of a modest number of antennas at the BS and their required DACs were not an issue in terms of energy consumption. This is due to the fact that DACs consume less energy than ADCs. Despite the diversity of research about DACs, very few allow the calculation of the increment of chip power dissipation as bit resolution increases bit-by-bit for a given technology. In order to roughly compare the consumption of both equipments, we make use of Table \ref{table1}, {which contains, in black,} the fabrication parameters for GaAs  4-bit  Analog-to-Digital Converter (AD) and  5-bit Digital-to-Analog (DA) converters, using a 0.7-$\mu$m MESFET self-aligned gate process
\cite{Naber} and the expression proposed in \cite{Orhan}. 
\begin{table}[htb!]
	\centering
	\caption{ADC and DAC fabrication parameters with the same technology -inferred approximated data in red}
	\begin{tabular}{|c c c c |} 
		\hline
		 & Resolution  & Sampling Rate & Power dissipation \\		  [0.5ex]
		 & (bits)   & (GHz) &  (mW)  \\
		\hline
         {DAC} & {4} & 1 & {$\approx 42.5$} \\
		ADC & 4 & 1 & 140 \\
		
		\hline
		DAC & 5 & 1 & 85 \\
		{ADC} & {5} & 1 & {$\approx280$}	\\
		[0.5ex] 
		\hline
		{DAC} & {6} & 1 & {$\approx170$} \\
		{ADC} & {6} & 1 & {$\approx560$}	\\
		[0.5ex] 
		\hline
		{DAC} & {{12}} & 1 & {{$\approx10880$} }\\
		{ADC} & {{12}} & 1 & {{$\approx35840$}}	\\
		[0.5ex] 
		\hline
	\end{tabular}
	\label{table1}
\end{table}

We start with the  expression\cite{Orhan} which relates the power consumed by an ADC  to the resolution in bits as follows: 
\begin{equation}
\label{adc_power}
\mathrm{P_{ADC} (b)}= \mathrm{c\:\tau\: 2^{b}}
\end{equation} 
where $ b $ stands for the resolution in bits, $c$ is a constant and $\tau$ is the sampling rate. 
From \eqref{adc_power}, we obtain $\frac{\mathrm{P_{ADC}} (4)}{\mathrm{P_{ADC}} (5)}=\frac{1}{2}$, which allows us to estimate  $\frac{\mathrm{P_{DAC}} (5)}{\mathrm{P_{ADC}} (5)}= \frac{\mathrm{P_{DAC}} (5)}{2\mathrm{P_{ADC}} (4)}$. With the help of Table \ref{table1}, we obtain $\mathrm{P_{DAC}} (5)\approx 30\% \mathrm{P_{ADC}} (5)$.
So, the DAC consumes around 30 $\% $ of
the energy  of the ADC with fixed parameter. From the results obtained before, we can roughly estimate the economy in energy by assuming that similarly to ADC, DAC consumption doubles with every extra bit of resolution, i.e., of $\mathcal{O}\left(2^{b} \right) $. {Therefore, a decrease in 6 resolution bits, for instance from 12 to 6 bits, represents a 98.4$\% $ lower consumption. This reduction of DAC consumption motivates our study.} Following the above reasoning, we can add the estimated DAC data with $\approx$ to Table \ref{table1}. 

\subsection{Computational complexity}

This subsection is devoted to the comparison of the computational complexity of the proposed and existing precoders in terms of required floating point operations (\text{FLOPs}). For this purpose, we will make use of the big O notation, i.e.,  $\mathcal{O}\left( \cdot\right)$. The number of \text{FLOPs} required by conventional ${\scriptstyle\mathrm{BD}}$ and ${\scriptstyle\mathrm{RBD}}$ algorithms are dominated by two \text{SVDs} \cite{Spencer1}. Since our system model is dedicated to broadcast channels, we can assume the widespread ratios $N_{b}\gg N_{u} \gg N_{j} $ and one of their resulting approximations $N_{u}- N_{j} \approx N_{u} $ to simplify the resulting expressions. Table \ref{table2} illustrates the computational cost required by the proposed ${\scriptstyle\mathrm{CQA}}$ and existing precoders.

\begin{table}[htp]\scriptsize
\caption{Computational complexity of proposed CQA and existing precoding algorithms}
\centering 
{\begin{tabular}{l l }
\hline\hline
Precoder & Computational cost (FLOPs) under $N_{b}\gg N_{u} \gg N_{j} $\\
\hline 
\\
ZF & $\frac{N_{b}^3}{2}+N_{b}^2\left(4 N_{u}-\frac{3}{2}\right)-N_{b}\left(N_{u}\right) $   \\ 
MMSE&  $\frac{N_{b}^3}{2}+N_{b}^2\left(4 N_{u}-\frac{3}{2}\right)-N_{b}\left(N_{u}-2\right) $   \\ 
BD  &  ${N_{b}^2}\left(32N_{j}+8\right) +N_{b}\left(32 N_{u}^{2} +72N_{j}^{2}\right)+ 64N_{u}^{2} $ \\ 
RBD  &  ${N_{b}^2}\left(32N_{j}+8\right) +N_{b}\left(32 N_{u}^{2} +72N_{j}^{2}\right)+ 64N_{u}^{2} $ \\
Bussgang ZF & $ \frac{N_{b}^3}{2}+N_{b}^2\left(4 N_{u}-\frac{3}{2}\right)-N_{b}\left(N_{u}\right) +C_{\delta}$   \\ 
Bussgang MMSE &$\frac{N_{b}^3}{2}+N_{b}^2\left(4 N_{u}-\frac{3}{2}\right)-N_{b}\left(N_{u}-2\right)+C_{\delta} $   \\
Proposed  & $ {N_{b}^2}\left(32N_{j}+8\right) +N_{b}\left(32 N_{u}^{2} +72N_{j}^{2}\right)+ 64N_{u}^{2} +C_{\delta}$ \\
CQA-BD &\\
Proposed  & $ {N_{b}^2}\left(32N_{j}+8\right) +N_{b}\left(32 N_{u}^{2} +72N_{j}^{2}\right)+ 64N_{u}^{2}+C_{\delta} $ \\
CQA-RBD &\\
\hline 
\end{tabular}}
\label{table2}
\end{table}
The extra cost $C_{\delta}$ required to convert ${\scriptstyle\mathrm{ZF}}$, ${\scriptstyle\mathrm{MMSE}}$, ${\scriptstyle\mathrm{BD}}$ and  ${\scriptstyle\mathrm{RBD}}$ into their corresponding  Bussgang-based precoders, which are listed in Table \ref{table2}, do not have significant impact on the total computational cost of their respective Bussgang-based algorithms. Due to their design, existing waterfilling and the proposed CQA-MAAS power allocation have a similar computational cost of $\mathcal{O}\left( N_{u}\right)$, which in practice does not result in significant additional cost to be imposed on ${\scriptstyle\mathrm{BD}}$ and ${\scriptstyle\mathrm{RBD}}$ to obtain their respective ${\scriptstyle\mathrm{CQA-BD-MAAS}}$ and ${\scriptstyle\mathrm{CQA-RBD-MAAS}}$ schemes. Table \ref{table3} depicts the complexity of the proposed ${\scriptstyle\mathrm{CQA-MAAS}}$ technique and existing WF power allocation. {A key advantage of BD-type precoders like the proposed ${\scriptstyle\mathrm{CQA-BD}}$ and ${\scriptstyle\mathrm{CQA-RBD}}$ algorithms over ZF and MMSE techniques is that due to their required SVDs they originate a power loading matrix that can be readily adjusted by power allocation algorithms. Therefore, the SVDs cost can be associated with the the precoders, whereas the power allocation algorithms adjust the power loading matrix resulting from the SVDs and require a reduced computational cost, $O(N_u)$.}

\begin{table}[htp]\small
\caption{Computational complexity of proposed CQA-MAAS and existing WF algorithms.}
\centering 
{\begin{tabular}{l c } 
\hline\hline 
 Technique & Computational cost (FLOPs) \\
\hline 
Waterfilling (WF) & $\mathcal{O}\left( N_{u}\right) $   \\ 
MAAS&  $\mathcal{O}\left( N_{u}\right) $   \\ 
\hline 
\end{tabular}}
\label{table3}
\end{table}

%
\section{Numerical results} 
\label{numerical_results}
\begin{figure}[htb!]
	\centering 
	\includegraphics[width=8.5cm,height=6.7cm]{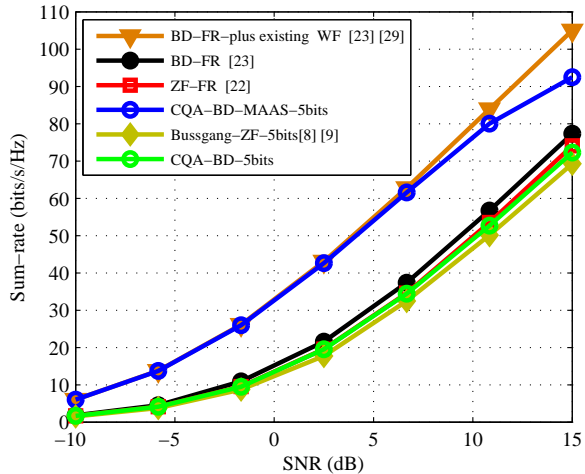}
	\caption{Achievable rates for CQA-BD and CQA-BD-MAAS, $5$quantization bits, via Bussgang theorem, Gaussian signals, compared to ZF-FR, BD-FR and BD-FR-plus existing WF. MU-MIMO configuration: $ N_{b}=64 $  and $ N_{u}=8 \times 2 $. }
	\label{R_ant_16_T_ant_64_b_5_200_4}
\end{figure}

\begin{figure}[htb!]
	\centering 
	\includegraphics[width=8.5cm,height=6.7cm]{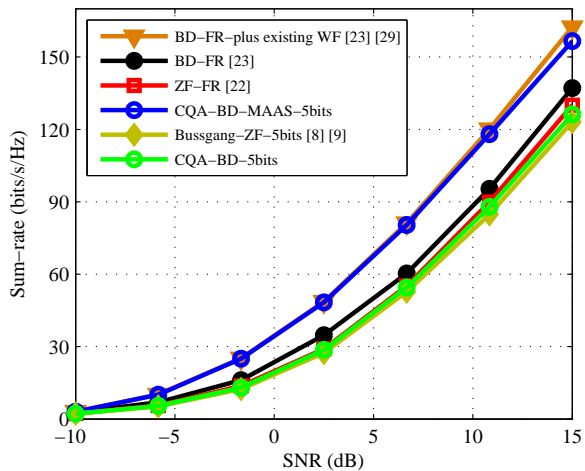}
	\caption{Achievable rates for CQA-BD and CQA-BD-MAAS, $5$quantization bits, via Bussgang theorem, Gaussian signals, compared to ZF-FR, BD-FR and BD-FR-plus existing WF. MU-MIMO configuration: $ N_{b}=64 $  and $ N_{u}=16 \times 2 $.  }
	\label{R_ant_32_T_ant_64_b_5_200_4}
\end{figure}

\begin{figure}[htb!]
	\centering 
	\includegraphics[width=8.5cm,height=6.7cm]{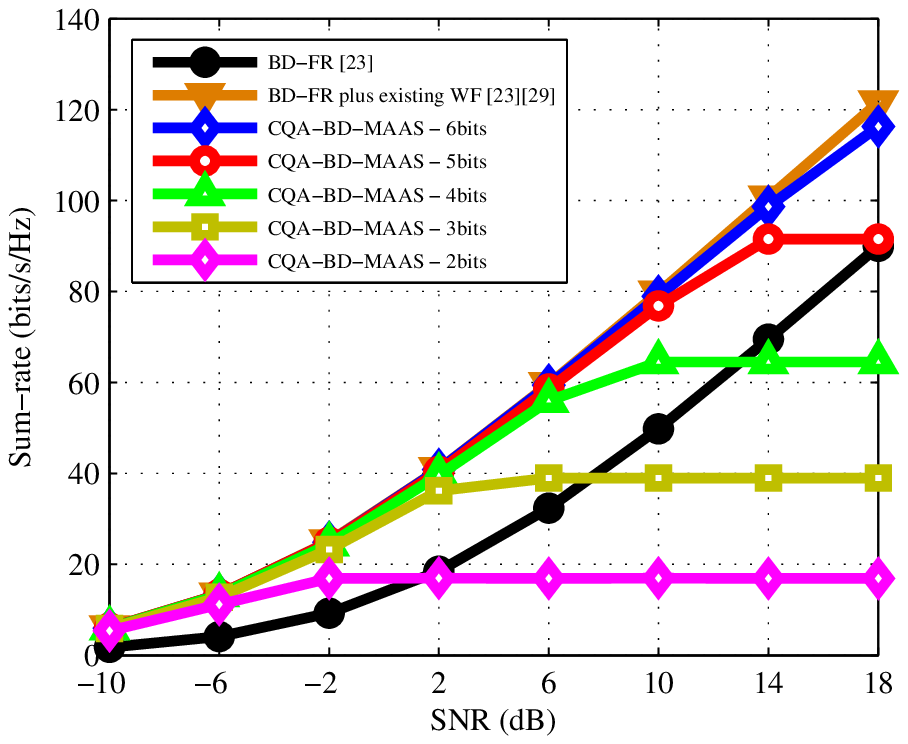}
	\caption{{Achievable rates for CQA-BD-MAAS, $2,3,4,5,6$ quantization bits, via Bussgang theorem, Gaussian signals, compared to BD-FR and BD-FR-plus existing WF. MU-MIMO configuration: $ N_{b}=64 $  and $ N_{u}=8 \times 2 $. }}
	\label{R_ant_16_T_ant_64_b_2_3_4_5_6_200_2}
\end{figure}

{In this section, we evaluate the performance of the proposed ${\scriptstyle\mathrm{CQA}}$ precoding techniques and the ${\scriptstyle\mathrm{CQA-MAAS}}$ power allocation strategy against the existing ZF, MMSE, ${\scriptstyle\mathrm{BD}}$ and ${\scriptstyle\mathrm{RBD}}$ precoders with full resolution and the Bussgang ZF and MMSE precoders \cite{Sven1,Sven2} with coarsely-quantized signals using simulations. We remark that we have only considered narrowband systems with flat fading channels, whereas the work in \cite{Sven2} considered frequency-selective channels with a multicarrier MIMO setting that employs Bussgang ZF and MMSE precoders per subcarrier. These precoders when applied per subcarrier in a multicarrier MIMO system are equivalent to the same precoders used for narrowband MIMO systems with flat fading channels. We also consider the influence of imperfect channel knowledge and spatial correlation on the sum-rates of our proposed algorithms.}
Although the acronyms employed in the figures have already been defined, for clarity, we list them here along with short explanations: 
\begin{enumerate}
    \item ${\scriptstyle\mathrm{CQA-MAAS}}$: proposed power allocation strategy with coarsely quantized signals.
    \item ${\scriptstyle\mathrm{BD-FR}}$: unquantized block-diagonalization algorithm that employs full resolution.
    \item ${\scriptstyle\mathrm{ZF-FR}}$: unquantized zero-forcing algorithm that employs full resolution.
    \item ${\scriptstyle\mathrm{BD-FR\: plus\: WF}}$: unquantized block-diagonalization algorithm that employs full resolution plus existing ${\scriptstyle\mathrm{WF}}$ \cite{Cover,Telatar,Paulraj}.
    \item ${\scriptstyle\mathrm{CQA-BD}}$: proposed block-diagonalization algorithm with coarsely quantized signals without ${\scriptstyle\mathrm{MAAS}}$.
    \item ${\scriptstyle\mathrm{CQA-BD-MAAS}}$: proposed  block-diagonalization algorithm with coarsely quantized signals with ${\scriptstyle\mathrm{MAAS}}$.  
    \item {${\scriptstyle\mathrm{PCH\: and\: ICH}}$: perfect and imperfect channel knowledge, respectively.}
 \end{enumerate}  
%
  We focus on two scenarios, whose MU-MIMO configurations are  $ \left(\mathit{N}_{b}=64, \mathit{N}_{u}=8 \times 2 \right)$ and  $ \left(\mathit{N}_{b}=64,  \mathit{N}_{u}=16 \times 2 \right)$  respectively. We model the channel matrix $\mathbf{H}_{j}$ of the $j$th user with entries given by complex Gaussian random variables with zero mean and unit variance. Additionally, it is assumed for simplicity that the channel is static during the transmission of each packet and that the antennas are uncorrelated. The channel is first considered perfectly known to the transmitter in the case of ${\scriptstyle\mathrm{CQA-BD}}$ and to both receiver and transmitter when using ${\scriptstyle\mathrm{CQA-MAAS}}$. {We set the number of independent trials to $5\times 10^2$ and the number of channels to $ 10^2$ symbols.}
%
Fig.\ref{R_ant_16_T_ant_64_b_5_200_4} depicts the sum-rates of the proposed ${\scriptstyle\mathrm{CQA-BD}}$ algorithm, and its variant equipped with ${\scriptstyle\mathrm{CQA-MAAS}}$, i.e., ${\scriptstyle\mathrm{CQA-BD-5bits}}$ and ${\scriptstyle\mathrm{CQA-BD-MAAS-5bits}}$, respectively, according to the first scenario, i.e, $ \left(\mathit{N}_{b}=64,  \mathit{N}_{u}=8 \times 2 \right)$. We  recall that ${\scriptstyle\mathrm{CQA-BD-MAAS-5bits}}$ is based on an approximation \eqref{Newman_3} obtained from a Neumann's truncated series. For comparison, we have also included the sum-rates of the $\scriptstyle\mathrm{BD-FR-plus\:existing\; WF}$, which in this specific study, can be considered an upperbound for $\scriptstyle \mathrm{BD}$-type algorithms, and the existing $\scriptstyle\mathrm{BD-FR}$ and ${\scriptstyle\mathrm{\mathrm{ZF-FR}}}$ precoding techniques, both under full resolution. It can be noticed that the following performance hierarchy is preserved over the considered range: ${\scriptstyle\mathrm{BD-FR-plus\:existing\; WF}\geq }$
 ${\scriptstyle\mathrm{CQA-BD-MASS-5bits} \geq }$ ${\scriptstyle\mathrm{BD-FR} \geq}$ ${\scriptstyle\mathrm{ZF-FR} \geq}$ ${\scriptstyle\mathrm{CQA-BD-5bits}\geq }$ ${\scriptstyle\mathrm{Bussgang\;ZF-5bits}}$.
 We can also  observe  the significant gap  between ${\scriptstyle\mathrm{CQA-BD-5bits}}$ and ${\scriptstyle\mathrm{CQA-BD-MASS-5bits}}$, resulting  from the incorporation  of ${\scriptstyle\mathrm{CQA-MASS}}$ into ${\scriptstyle\mathrm{CQA-BD-5bits}}$. This gain in terms of achievable sum-rate, which is
  obtained for low SNR, remains highly satisfactory over the  remaining range of values. We also notice that the proposed ${\scriptstyle\mathrm{CQA-BD}}$ algorithm outperforms the  ${\scriptstyle\mathrm{Bussgang-ZF}}$ algorithm \cite{Pinto} by up to $\mathrm{3dB}$  for the same performance. Additionally,  ${\scriptstyle\mathrm{CQA-BD-MASS-5bits}}$ and ${\scriptstyle\mathrm{BD-FR-plus\:existing\; WF}}$ are very close in the range $\mathrm{\left[-10\quad 7\right)dB}$, whereas in the range $\mathrm{\left[7\quad 15\right]dB}$ the gap between them is very small. This gap decreases in a scenario composed of more receive antennas and also in the event of more quantization bits.
\newline
Fig.\ref{R_ant_32_T_ant_64_b_5_200_4} illustrates the sum-rate performance in the second configuration mentioned before, i.e$ \left(\mathit{N}_{b}=64,  \mathit{N}_{u}=16 \times 2 \right)$. In this setting, in which we maintain the levels of quantization, in the range $\mathrm{\left[-1.6\quad 15\right]dB}$ all curves displayed move up, reaching higher sum-rates. Their ranking, which is defined by the inequalities commented on previous figure remains the same, however, considering from the bottom to the top, the first group of curves composed of ${\scriptstyle\mathrm{Bussgang\;ZF-5bits}}$, ${\scriptstyle\mathrm{CQA-BD-5bits}}$, ${\scriptstyle\mathrm{ZF-FR}}$ and ${\scriptstyle\mathrm{BD-FR}}$ are closer to the first one, comprised by ${\scriptstyle\mathrm{CQA-BD-MASS-5bits}}$ and ${\scriptstyle\mathrm{BD-FR-MAAS}}$. It can also be noticed that the two last mentioned curves, which in the prior figure, were already very close in the range  $\mathrm{\left[-10\quad 6.6\right)dB}$, now become  closer from $\mathrm{6.6dB}$ to $\mathrm{15dB}$. 
\newline
In Fig.\ref{R_ant_16_T_ant_64_b_2_3_4_5_6_200_2}, which corresponds to the first scenario, i.e.,$ \left(\mathit{N}_{b}=64,  \mathit{N}_{u}=8 \times 2 \right)$, it is shown the effect of the increase in the number of quantization bits on the performance of the proposed ${\scriptstyle\mathrm{CQA-BD-MASS}}$ in the considered range of $\mathrm{SNR}$. In order to better assess its effectiveness, we have also plotted ${\scriptstyle\mathrm{BD-FR}}$.   ${\scriptstyle\mathrm{CQA-BD-MASS-6bits}} $ clearly works as well as its upperbound ${\scriptstyle\mathrm{BD-FR-plus\:existing\; WF}}$.

\begin{figure}[htb!]
	\centering 
	\includegraphics[width=8.5cm,height=6.7cm]{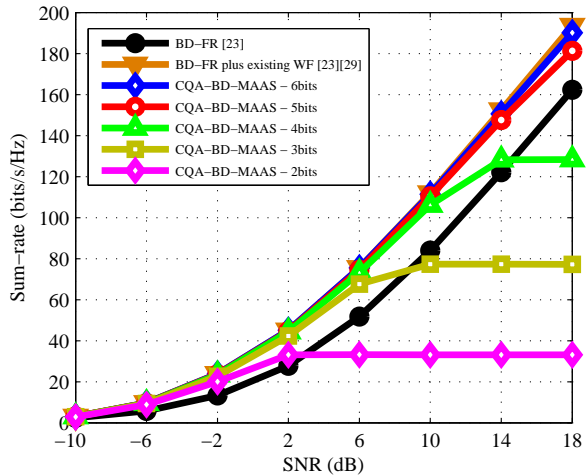}
	\caption{{Achievable rates for CQA-BD-MAAS, $2,3,4,5,6$ quantization bits, via Bussgang theorem, Gaussian signals, compared to that reached by BD-FR and { BD-FR-plus existing WF}. MU-MIMO configuration: $ N_{b}=64 $  and $ N_{u}=16 \times 2 $. } }
	\label{R_ant_32_T_ant_64_b_2_3_4_5_6_2}.
\end{figure}

In Fig.\ref{R_ant_32_T_ant_64_b_2_3_4_5_6_2}, we assess the degree in which the sum-rates increase when the level of quantization varies in the second configuration,i.e., $ \left(\mathit{N}_{b}=64, \mathit{N}_{u}=16 \times 2 \right)$. Both ${\scriptstyle\mathrm{CQA-BD-MASS-6bits}} $ and  ${\scriptstyle\mathrm{CQA-BD-MASS-5bits}} $ are closer to ${\scriptstyle\mathrm{BD-FR-MAAS}}$. The performance of ${\scriptstyle\mathrm{CQA-BD-MASS}}$  achieved in this scenario under $5$ and $6$bits quantization cannot be only justified by the increase in the levels of quantization, but also by the extra receive antennas. It can be noticed that  the size of the coefficient $\epsilon$ defined in \eqref{coef_epsilon} is a condition of the approximation \eqref{achievable_sum_rate_BD_power_alloc_1} provided by Theorem \ref{theorem2}. In other words, a decrease in $\epsilon$ resulting from an increased number of receive antennas leads to more accurate approximations of the sum-rate computed by \eqref{achievable_sum_rate_BD_power_alloc_1}. 

\vspace{-1.0em}
\begin{figure}[htb!]
	\centering 
	\includegraphics[width=8.5cm,height=6.7cm]{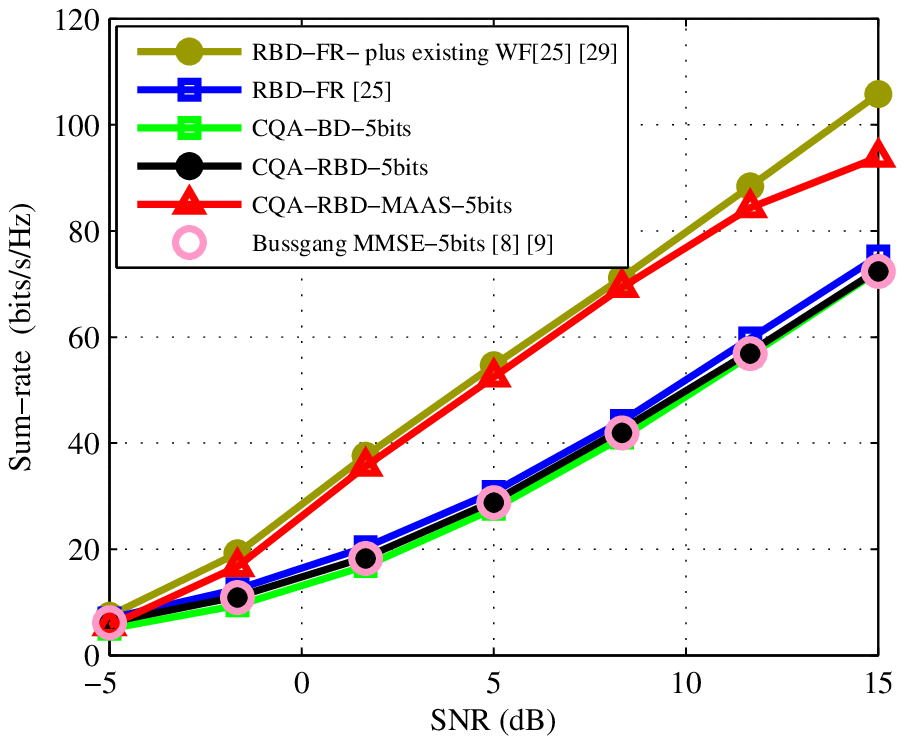}
	\caption{{Achievable rate for  ${\scriptstyle\mathrm{CQA-RBD-MAAS-5bits}}$, via Bussgang theorem, Gaussian signals, compared to that reached by ${\scriptstyle\mathrm{CQA-RBD-5bits}}$, ${\scriptstyle\mathrm{CQA-BD-5bits}}$, ${\scriptstyle\mathrm{RBD-FR}}$ and ${\scriptstyle\mathrm{RBD-FR}}$-plus existing ${\scriptstyle\mathrm{WF}}$. MU-MIMO configuration: $ N_{b}=64 $  and $ N_{u}=8 \times 2 $.  }}
	\label{Rbd_behav_WF_MAAS_5b_64_16_500_1}.
\end{figure}

\vspace{4mm}
 Fig.\ref{Rbd_behav_WF_MAAS_5b_64_16_500_1} illustrates the closeness between  ${\scriptstyle\mathrm{CQA-RBD-MAAS-5bits}}$
 and its theoretical upper bound, ${\scriptstyle\mathrm{RBD-FR-plus\:existing\; WF}}$ and the gain that $\scriptstyle\mathrm{CQA-MAAS}$ incorporates into 
 $\scriptstyle\mathrm{CQA-RBD-5bits}$ to convert it into $\scriptstyle\mathrm{CQA-RBD-MAAS-5bits}$. We have  also included $\scriptstyle \mathrm{RBD-FR}$, $\scriptstyle\mathrm{CQA-BD-5bits} $ and ${\scriptstyle\mathrm{Bussgang\;MMSE-5bits}}$ for comparisons. We can observe that, except for the neighborhood of $\mathrm{-5dB }$, the following non strict inequalities are preserved over the considered range in a similar way to that in Fig.\ref{R_ant_16_T_ant_64_b_5_200_4}: $\scriptstyle\mathrm{RBD-FR-plus\:existing WF}$ 
 $\scriptstyle\mathrm{CQA-RBD-MASS-5bits} \geq $ ${\scriptstyle\mathrm{RBD-FR} \geq}$ ${\scriptstyle\mathrm{CQA-RBD-5bits} \geq}$ ${\scriptstyle\mathrm{Bussgang\;MMSE-5bits} \geq}$ $\scriptstyle\mathrm{CQA-BD-5bits} $. The large gap between $\scriptstyle\mathrm{CQA-RBD-MAAS-5bits}$ and $\scriptstyle \mathrm{CQA-RBD-5bits}$ in the range $\mathrm{\left[-1.6\quad 15\right]dB}$ and also its closeness to its theoretical upperbound, i.e., $\scriptstyle \mathrm{RBD-FR}$ plus existing $\scriptstyle \mathrm{WF}$, make clear the effectiveness of the proposed $\scriptstyle \mathrm{CQA-MAAS}$ power allocation combined with the Bussgang theorem for quantized signals applied to $\scriptstyle \mathrm{RBD}$ algorithms.

\begin{figure}[htb!]
	\centering 
	\includegraphics[width=8.5cm,height=6.7cm]{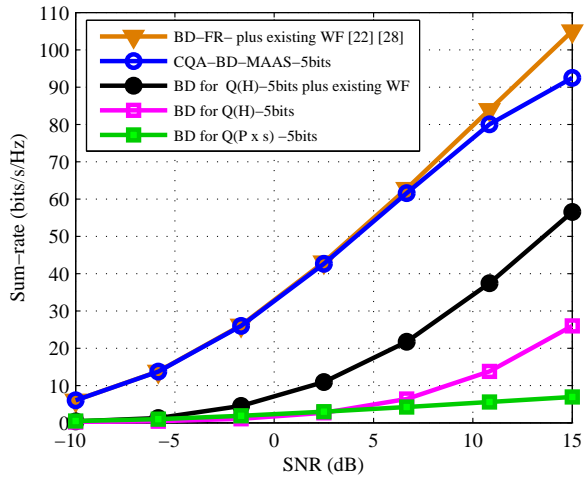}
	\caption{{Achievable rate for  CQA-BD-MAAS-5bits,via Bussgang theorem, Gaussian signals, compared to that provided by  BD-FR-plus existing WF, BD for 5bit-quantized channel, BD for 5bit-quantized precoded signal and BD for 5bit-quantized channel plus existing WF. MU-MIMO configuration: $ N_{b}=64 $  and $ N_{u}=8 \times 2 $.  }}
	\label{BD_effects_quant_5b_64_16_500}.
	\end{figure}

In the example shown in Fig.\ref{BD_effects_quant_5b_64_16_500}, we compare the proposed ${\scriptstyle\mathrm{CQA-BD-MAAS-5bits}}$  to variations of  5bit-quantized $\scriptstyle \mathrm{BD}$ algorithms, which  employ traditional quantization, i.e., they are not based on Busgang theorem. It is also plotted   ${\scriptstyle\mathrm{BD-FR-plus\:existing\: WF}}$ as their theoretical upperbound. We have discarded a possible curve representing ${\scriptstyle\mathrm{BD\: for\:Q\left(\mathbf{P}\times\mathbf{s}\right)-5bits\:plus\: existing\:WF}}$ due to its poor performance. The closeness of the proposed ${\scriptstyle\mathrm{CQA-RBD-MAAS-5bits}}$ to its upperbound ${\scriptstyle\mathrm{BD-FR-plus\:existing\; WF} }$ and the huge gap between it and the ${\scriptstyle\mathrm{BD\: for\:Q\left(\mathbf{H}\right)-5bits\:plus\: existing\:WF}}$ in all considered range make clear its impressive performance. 
%

{We now consider the impact of practical aspects. Specifically, we have considered the model for imperfect channel knowledge and spatial correlation $\hat{\mathbf H}=\mathbf {H}\;\tilde{\mathbf{R}}\ ^{\frac{1}{2}} + \mathbf {E}$, where $\tilde{\mathbf{R}}$ represents the complex transmit correlation matrix \cite{Loyka,Zu} whose elements are
\begin{align}
\tilde{R}_{ij}=\left\{\begin{array}{ll} r^{j-i},& i\leq j\\ r_{ji}^*,& i>j \end{array} \right.,|r|\leq 1
\end{align}
where $\lvert\mathrm{r}\rvert <1$. It can be noticed that the absolute values of the entries $\lvert \tilde{\mathrm{R}}_{\left(i,j\right)} \rvert$ corresponding to the closest antennas are larger than the others. The error matrix  $\mathbf{E}$ is modeled \cite{Zu} as a complex Gaussian noise with i.i.d entries of zero mean and variance $\sigma_{e}^{2}$. In our next examples, we have employed large values of correlations between the neighboring antennas, i.e., $\lvert\mathrm{r}\rvert=0.72$ and $ 0.91 $, respectively. The variance  $\sigma_{e}^{2}$ of the  feedback error matrix  $\mathbf{E}$ has been set to $0.16$.} 

\begin{figure}[htb!]
	\centering 
	\includegraphics[width=8.5cm,height=6.7cm]{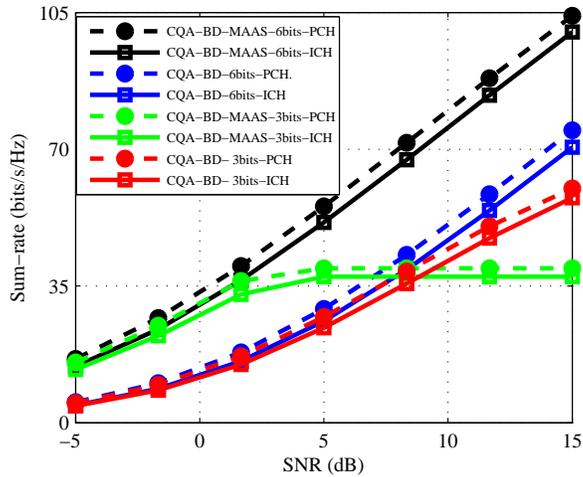}
	\caption{{{Achievable rates for CQA-BD and  CQA-BD-MAAS for 6 and 3-bit quantization under perfect and imperfect channel knowledge. MU-MIMO configuration: $ N_{b}=64 $  and $ N_{u}=8 \times 2 $.
	 {$\lvert\mathrm{r}\rvert=0.72$  and $\sigma_{e}^{2}=0.16$.} }}}
	\label{Channel_robustness_BD_3b_6b_v1}.
	\end{figure}

{In Fig. \ref{Channel_robustness_BD_3b_6b_v1}, we assess the performance of ${\scriptstyle\mathrm{CQA-BD}}$ and ${\scriptstyle\mathrm{CQA-BD-MAAS}}$ in the presence of imperfect channel knowledge and spatial correlation using $3$ and $6$ bits. The results show that the impact of imperfect channel knowledge is not significant in terms of performance degradation of the precoders. However, the performance degradation of ${\scriptstyle\mathrm{CQA-BD-MAAS}}$ can become significant for $3$ bits}. 

\begin{figure}[htb!]
	\centering 
	\includegraphics[width=8.5cm,height=6.7cm]{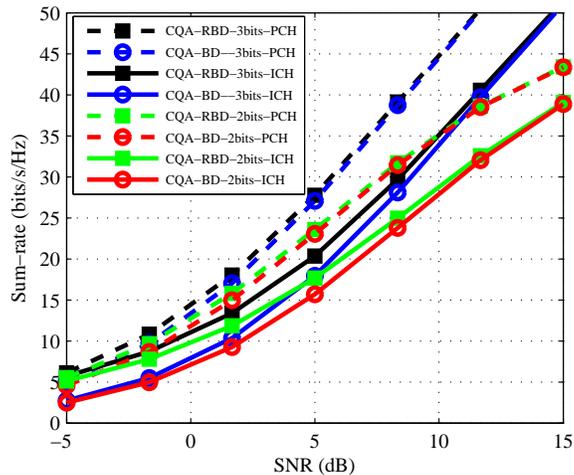}
	\caption{{{Achievable rates for CQA-RBD and  CQA-BD for 2 and 3-bit quantization, under perfect and imperfect channel knowledge. MU-MIMO configuration: $ N_{b}=64 $  and $ N_{u}=8 \times 2 $. {$\lvert\mathrm{r}\rvert=0.91$  and $\sigma_{e}^{2}=0.16$.} }}}
	\label{Channel_robustness__RBD_BD_2b_3b_correl_ant_091_var_err_016_v2}.
	\end{figure}

{In Fig. \ref{Channel_robustness__RBD_BD_2b_3b_correl_ant_091_var_err_016_v2}, we assess the performance of ${\scriptstyle\mathrm{CQA-BD}}$ and ${\scriptstyle\mathrm{CQA-RBD}}$ in the presence of imperfect channel knowledge and spatial correlation using $2$ and $3$ bits. The results show that ${\scriptstyle\mathrm{CQA-RBD}}$ outperforms ${\scriptstyle\mathrm{CQA-BD}}$ and the advantage of ${\scriptstyle\mathrm{CQA-RBD}}$ in performance is more pronounced for scenarios with imperfect channel knowledge, which indicates the increased robustness of ${\scriptstyle\mathrm{CQA-RBD}}$. Specifically, the sum-rate performance of ${\scriptstyle\mathrm{CQA-RBD}}$ is up to 30$\%$ higher than that of ${\scriptstyle\mathrm{CQA-BD}}$ in scenarios with imperfect channel knowledge.}

\section{Conclusions}
\label{conclusions}

{We have proposed the ${\scriptstyle\mathrm{CQA-BD}}$ and ${\scriptstyle\mathrm{CQA-RBD}}$ precoding and the ${\scriptstyle\mathrm{CQA-RBD-MAAS}}$ power allocation algorithms for massive MIMO systems that employ coarse quantization using DACs with few bits. The proposed ${\scriptstyle\mathrm{CQA-BD}}$ and ${\scriptstyle\mathrm{CQA-RBD}}$ precoding algorithms outperform existing Bussgang-ZF and Bussgang-MMSE precoders for systems with coarse quantization, resulting in sum-rate gains of up to $15$\% while requiring a comparable computational cost. Moroever, ${\scriptstyle\mathrm{CQA-RBD-MAAS}}$ can obtain gains in sum-rate of up to $30$ \% over schemes without power allocation and comparable performance to full-resolution schemes with precoding and WF power allocation. These findings include scenarios with perfect and imperfect channel knowledge as well as spatial correction between antennas. Finally, the proposed algorithms can be used in massive MIMO systems and contribute to substantial reduction in power consumption. }

\appendix
Here we provide the derivation of \eqref{achievable_sum_rate_BD_RBD}.
\subsection{Assumptions}
According to Subsection \ref{sysmodel}, we assume that the cross covariance matrices $\mathbf{R}_{sf}=\mathbb E[\mathbf{s}\mathbf{f}^{H}]$ and $\mathbf{R}_{sn}=\mathbb E[\mathbf{s}\mathbf{n}^{H}]$ and $\mathbf{R}_{fn}=\mathbb E[\mathbf{f}\mathbf{n}^{H}]$ are all equal to zero and so are $\mathbf{R}_{fs}$, $\mathbf{R}_{ns}$ and $\mathbf{R}_{nf}$.
Additionally, the distortion vector is assumed to be Gaussian.
 \subsection{Development}
 We start by combining \eqref{downlink_channel_model} with \eqref{bus_theo1}, obtaining:
\begin{equation}
\label{downlink_channel_model_mod_Paulraj}
\mathbf{y}= \mathbf{HTPs}+\mathbf{Hf}\:+\:\mathbf{n},
\end{equation}
where we can define a distortion-plus-noise vector
\begin{equation}
\label{distortion_plus_noise_vector}
\mathbf{\tilde{n}}= \mathbf{Hf}\:+\:\mathbf{n}.
\end{equation}
We can estimate the correlation matrix $\mathbf{R}_{sqsq}$ of the quantized vector \eqref{bus_theo1}, as follows: 
\begin{align}
\label{autocorrel_quantized_vector}
\mathbf{R}_{sqsq}&=\mathbb E[\mathbf{(TPs+f)}\mathbf{(TPs+f)}^H]\nonumber\\&=  \mathbb E[\mathbf{TPs}\mathbf{s}^{H}\mathbf{P}^{H}\mathbf{T}^{H}+\mathbf{f}\mathbf{f}^{H}]\nonumber\\&=\delta^{2}\sigma^{2} \mathbf{P}\mathbf{P}^{H} +\mathbf{R}_{ff},
\end{align}
where we made use of \eqref{approx_diag_mat} and the autocorrelation matrix of the symbol vector  $\mathbf{R}_{ss}=\mathbb E[\mathbf{s}\mathbf{s}^{H}]=\sigma_{s}^{2}\mathbf{I}_{Nu}$, in which  $ \sigma_{s}^{2} $ is its variance. The term  $\mathbf{R}_{ff}=\mathbb E[\mathbf{f}\mathbf{f}^{H}]$ stands for the autocorrelation of the distortion vector $\mathbf{f}$.
Next, we can notice that in full resolution, since there is no quantization and its associated distortion, \eqref{bus_theo1} turns into
\begin{equation}
\label{bus_theo1_reduced}
\ \mathbf{s}_{q}=\mathbf{Ps},
\end{equation}
where we make $\mathbf{T}_{n,n}= \mathbf{I}_{Nb\times Nb}$, i.e., $\delta= 1$  in \eqref{approx_diag_mat}, and assume that $\mathbf{f}=\textbf{0}_{Nb}$. 
Now, we calculate the autocorrelation of the full resolution precoded symbol vector \eqref{bus_theo1_reduced} as follows:
\begin{align}
\label{autocorrel_fr_precod_symbol_vector}
\mathbf{R}_{sqsq}&=\mathbb E[\mathbf{(Ps)}\mathbf{(Ps)}^H]\nonumber\\&=  \mathbb E[\mathbf{Ps}\mathbf{s}^{H}\mathbf{P}^{H}]\nonumber\\&=\sigma_{s}^{2} \mathbf{M}\mathbf{P}^{H},
\end{align}
By equating \eqref{autocorrel_quantized_vector} and \eqref{autocorrel_fr_precod_symbol_vector}, we can obtain the expression of the autocorrelation of the distortion vector: 
\begin{align}
\label{autocorr_distort}
&\delta^{2}\sigma_{s}^{2} \mathbf{P}\mathbf{P}^{H} +\mathbf{R}_{ff}=\sigma_{s}^{2} \mathbf{P}\mathbf{P}^{H}\nonumber\\&\therefore \mathbf{R}_{ff}=\left(1- \delta^{2}\right)\sigma_{s}^{2}\mathbf{P}\mathbf{P}^{H} 
\end{align}
We can then compute the autocorrelation matrix of \eqref{downlink_channel_model_mod_Paulraj}:
\begin{align}
\label{cov_matrix_channel_model_mod_Paulraj}
\mathbf{R}_{yy}&=\mathbb E[\mathbf{y}\mathbf{y}^H] \nonumber\\&= \left( \mathbf{HTP}\right) \:\mathbf{R}_{ss}\left( \mathbf{HTP}\right) ^{H}\nonumber\\&+\mathbf{H}\:\mathbf{R}_{ff}\mathbf{H}^{H}\:+\:\mathbf{R}_{nn},
\end{align}
where $\mathbf{R}_{ss}=\mathbb E[\mathbf{s}\mathbf{s}^{H}]$,  $\mathbf{R}_{ff}=\mathbb E[\mathbf{f}\mathbf{f}^{H}]$ and $\mathbf{R}_{nn}=\mathbb E[\mathbf{n}\mathbf{n}^{H}]$ are the autocorrelation matrices of the signal,  the distortion and the noise vectors, respectively. 
Similar procedure applied to the distortion-plus-noise vector \eqref{distortion_plus_noise_vector}, considering the conditions above, yields
\begin{align}
\label{cov_matrix_dist_plus_noise}
\mathbf{R}_{\tilde{n}\tilde{n}}=\mathbb E[{\tilde{\mathbf{n}}\tilde{\mathbf{n}}^H}]=\mathbf{H}\mathbf{R}_{ff}\mathbf{H}_{H} +\mathbf{R}_{nn},
\end{align}
From the principles of information theory \cite{Cover} and the capacity of a  frequency flat deterministic MIMO channel \cite{Paulraj}, we can bound  the achievable rate in bits per channel use at which information can be sent with arbitrarily low probability of error by the mutual information of a Gaussian channel, i.e.
\begin{align}
\label{reachable_rate}
\mathit{C} \leqq \mathit{I}\left(\mathbf{s},\mathbf{y} \right)&= \Upsilon\left(\mathbf{y} \right)-\Upsilon\left(\mathbf{y}\rvert\mathbf{s} \right)\nonumber\\&=\Upsilon\left(\mathbf{y} \right)-\Upsilon\left(\mathbf{\tilde{n}}\right)\nonumber\\&= \log_{2}\left[ \det\left( \pi e\mathbf{R}_{yy}\right) \right] -\log_{2}\left[ \det\left( \pi e\mathbf{R}_{\tilde{n}\tilde{n}}\right) \right]\nonumber\\&= \log_{2}\left[ \det\left( \mathbf{R}_{yy}\right) \right] -\log_{2}\left[ \det\left(  \mathbf{R}_{\tilde{n}\tilde{n}}\right) \right]\nonumber\\&=
\log_{2}\left[ \det\left( \mathbf{R}_{yy}\mathbf{R}^{-1}_{\tilde{n}\tilde{n}}\right) \right] 
\end{align}
where $\Upsilon\left(\mathbf{y}\right) $  and $\Upsilon\left(\mathbf{y}\rvert\mathbf{s}_{q} \right)$ are the differential and the conditional differential entropies of  $\mathbf{y}$, respectively.
By combining \eqref{cov_matrix_channel_model_mod_Paulraj} and \eqref{cov_matrix_dist_plus_noise} with \eqref{reachable_rate}, we have:
\begin{align}
\label{reachable_rate_2}
\mathit{C} &\leqq\log_{2}\left\lbrace \det\left[ \left( \left( \mathbf{HTP}\right) \:\mathbf{R}_{ss}\left( \mathbf{HTP}\right) ^{H}\right)\right.\right.\nonumber\\&\left.\left.\left(\mathbf{H}\mathbf{R}_{ff}\mathbf{H}^{H} +\mathbf{R}_{nn} \right)^{-1} +\mathbf{I}_{N_{u}} 
\right] \right\rbrace   
\end{align}
In Section \ref{sysmodel}, we have defined the total power as  $\mathit{P}= \mathit{SNR}\ \mathrm{N}_{0}$. From the definition of the noise vector, also in that Section, we can express  its covariance matrix as $\mathbf{R}_{nn}=\mathrm{N}_{0}\mathbf{I}_{N_{u}} $ 
Combining the two previously mentioned expressions, we obtain
\begin{align}
\label{noise_covar_matrix}
\mathbf{R}_{nn}=\frac{\mathit{P}}{SNR}\;\mathbf{I}_{N_{u}}
\end{align}
Recalling from Section \ref{sysmodel} that $\mathbf{R}_{ss}\approx \mathbf{I}_{N_{u}} $ and  assuming that the total power  is given by $\mathit{P}= {\rm trace}\left( \mathbf{R}_{ss}\right)= N_{u}  $, \eqref{noise_covar_matrix} turns into
\begin{align}
\label{noise_covar_matrix_mod}
\mathbf{R}_{nn}=\frac{\mathit{N_{u}}}{SNR}\;\mathbf{I}_{N_{u}}
\end{align}
By combining \eqref{reachable_rate_2}, \eqref{autocorr_distort} \eqref{approx_diag_mat} and the expression of $\mathbf{R}_{ss}$ previously mentioned with \eqref{noise_covar_matrix_mod}, followed by algebraic manipulation, we obtain
\begin{align}
\label{achievable_sum_rate_BD_RBD_final}
\mathit{C}=& \log_{2}\left\lbrace\det\left[ \mathbf{I}_{Nu} + \delta^{2} \frac{\mathit{SNR}}{\mathit{N}_{u}}\mathbf{\left( HP\right) } \mathbf{\left( HP\right) }^{H}\right.\right.\nonumber\\&\left.\left. \left(\left(1-\delta^{2} \right)\frac{\mathit{SNR}}{\mathit{N}_{u}}\mathbf{\left( HP\right) } \mathbf{\left( HP\right) }^{H} +\mathbf{I}_{Nu}              \right)^{-1}                 \right]  \right\rbrace \qed
\end{align}

{\linespread{0.9}

}
\end{document}